\documentclass[12pt,preprint]{aastex}
\usepackage{natbib} 
\def\deg{^\circ}
\def\MS{M_\odot}
\def\ppar{p_\parallel}
\def\pper{p_\perp}
\def\pbig{P_\perp}
\def\shat{\hat{\bf s}}

\def\rsun{R_\odot}
\def\rmax{R_{max}}
\def\cpar{c_\parallel}
\def\cper{c_\perp}
\def\parper{{\parallel,\perp}}
\def\parper{{\parallel,\perp}}
\def\Tppar{T_{p\parallel}}
\def\Tpper{T_{p\perp}}
\def\rcusp{r_{\rm cusp}}
\def\dn{\delta\nu}

\shorttitle{Solar Minimum Corona}
\shortauthors{V\'asquez, van Ballegooijen, and Raymond}

\begin{document}

\title{The Effect of Proton Temperature Anisotropy on the Solar
Minimum Corona and Wind} 

\author{Alberto M. V\'asquez \altaffilmark{1}, Adriaan A. van
Ballegooijen, and John C. Raymond}
\affil{Harvard-Smithsonian Center for Astrophysics, 60 Garden Street,
MS 15, Cambridge, MA 02138, USA.}
\email{avasquez@cfa.harvard.edu}

\altaffiltext{1}
{Currently at the following postal address:\\
\emph{Instituto de Astronom\'\i a y F\'\i sica del Espacio},
CC 67 - Suc 28, (1428) Ciudad de Buenos Aires, Argentina.}

\begin{abstract}
A semi-empirical, axisymmetric model of the solar minimum corona is
developed by solving the equations for conservation of mass and
momentum with prescribed anisotropic temperature distributions. In the
high-latitude regions, the proton temperature anisotropy is strong and
the associated mirror force plays an important role in driving the
fast solar wind; the critical point where the outflow velocity equals
the parallel sound speed ($v = c_\parallel$) is reached already at 1.5
$\rsun$ from Sun center. The slow wind arises from a region with open
field lines and weak anisotropy surrounding the equatorial streamer
belt. The model parameters were chosen to reproduce the observed
latitudinal extent of the equatorial streamer in the corona and at
large distance from the Sun. We find that the magnetic cusp of the
closed-field streamer core lies at about 1.95 $\rsun$. The transition
from fast to slow wind is due to a decrease in temperature anisotropy
combined with the non-monotonic behavior of the non-radial expansion
factor in flow tubes that pass near the streamer cusp. In the
slow wind, the plasma $\beta$ is of order unity and the critical
point lies at about 5 $\rsun$, well beyond the magnetic cusp. The
predicted outflow velocities are consistent with $\rm O^{5+}$ Doppler
dimming measurements from UVCS/{\it SOHO}. We also find good agreement
with polarized brightness ($pB$) measurements from LASCO/{\it SOHO}
and H I Ly$\alpha$ images from UVCS/{\it SOHO}.
\end{abstract}


\section{Introduction}

At the time of cycle minimum, the solar corona is more or less
axisymmetric and stable for many months. The polar coronal holes
are separated by an equatorial streamer belt that encircles the Sun.
The fast solar wind originates from the coronal holes, while the
slow wind originates from the vicinity of the streamer belt.
The physical processes that drive the fast and slow wind are only
partially understood. In particular, it is unclear why the Sun has
such a bimodal outflow pattern with distinctly different physical
conditions in the fast and slow wind. Does this distinction between
fast and slow wind already exist at low heights in the corona, or
does it arise only at larger distance from the Sun? What causes
the transition from fast to slow wind as we approach the solar
equator? What is the role of open and closed magnetic fields in this
transition? To answer such questions, an empirical description of
the corona is needed; i.e., a description of temperature, density,
velocity and magnetic field as function of latitude, longitude and
radial distance from the Sun. Such modeling is a necessary step in
the study of the physical processes by which the fast and slow wind
are generated.

Axisymmetric, semi-empirical models of the solar corona have been
developed by many authors \citep[e.g.,][]{PK71,YP77,SS82,WY87,CO90,
CB93,WWSP93,WWSP98,lionello2001}. For example, \citet{sittler99} used
empirically derived electron density profiles to construct a model of
the magnetic field, outflow velocity, effective temperature, and
effective heat flux. These parameters are derived by solving the
equations for conservation of mass, momentum and energy, and the
magnetic induction equation. The model provides an estimate of the
large-scale surface magnetic field at the Sun, which is estimated to
be 12-15 G. The authors predict that the large-scale surface field is
dominated by an octupole term. More recently, three-dimensional models
of the corona have also been developed \citep[e.g.,][]{LH90,wang97a,
riley2001}.

In this paper we develop a coronal model that synthesizes
observational data obtained with instruments on the {\it Solar and
Heliospheric Observatory} ({\it SOHO}) satellite, in particular data
from the Ultraviolet Coronagraph Spectrometer \citep[UVCS;][]{kohl95}.
UVCS observations have shown that the minor ions in coronal holes have
kinetic temperatures that are much larger than those for protons and
electrons \citep{kohl98,cranmer99}. Here kinetic temperature refers
to the total velocity dispersion of the particles, including 
thermal and non-thermal components. Detailed analysis of kinetic
temperatures for different ions has shown that some ions have higher
temperature than others, indicating {\it preferential heating}
of certain ions. 
Furthermore, Doppler dimming analysis of spectral lines such as O VI
$\lambda\lambda$1032 and 1037 has shown that some minor ions have
{\it anisotropic velocity distributions}: the velocity dispersion
perpendicular to the (nearly radial) magnetic field in coronal holes
is significantly larger than that parallel to the field. This high
temperature and large temperature anisotropy of the ions is believed
to be caused by dissipation of transverse waves
\citep{dusen81,isen83,tu+marsch97,cranmer99,hu+habbal99}. A similar
but smaller anisotropy may also exist for the protons
\citep{cranmer99}. In the presence of a diverging magnetic field,
charged particles experience an outward force that causes the
perpendicular motion of the particles to be converted into parallel
motion.  Therefore, if the perpendicular energy of the protons is
continually replenished by wave dissipation, the protons can maintain
an anisotropic velocity distribution ($\Tpper > \Tppar$) and there is
a net outward force on the protons. This so-called mirror force plays
an important role in driving the fast solar wind \citep[see review
by][]{cranmer2002}.

The purpose of the present paper is to include the effects of proton
temperature anisotropy into a global model of the solar corona. The
paper is organized as follows. In \S 2 we discuss observations of
streamers and the relationship between the streamer and the slow solar
wind. In \S 3 we present observational constraints on our global model,
including temperature and density in the corona and magnetic flux
at the coronal base. In \S 4 we present a method for solving the
coronal force balance equations, taking into account the mirror force.
In \S 5 we present our model results for coronal magnetic structure
and outflow velocity. In \S 6 we derive images of visible light
polarization brightness and H I Ly$\alpha$ intensities from our model,
and we compare our results with observations. The main results of this
work are discussed in \S 7.

\section{Streamer Structure and the Origin of the Slow Wind}

The K corona is produced by Thomson scattering of photospheric white
light by free electrons in the corona. In a study of the solar minimum
corona using the Large Angle and Spectrometric Coronagraph (LASCO) on
{\it SOHO}, \citet{wang97a} found that the large-scale structure of
the coronal streamer belt at 3 $\rsun$ and beyond can be reproduced
with a model in which the scattering electrons are concentrated around
a single, warped current sheet that encircles the Sun. The angular
width of this sheet is only a few degrees. The bright, narrow spikes
seen in LASCO C2 and C3 images occur wherever the sheet is oriented
edge-on in the plane of the sky \citep{wang98}. In contrast, the
latitudinal extent of the slow wind, as determined from {\it in-situ}
measurements, is about $\pm 20\deg$ \citep{Su99},
much larger than the thickness of the coronal plasma sheet observed
with LASCO. This implies that the bulk of the slow wind originates
from open field lines outside the observed plasma sheet
\citep{wang98}.

\citet{sheeley97} used time-lapse sequences of LASCO images to track
the outward motion of small density enhancements (``blobs'') in the
plasma sheet. These blobs originate in the high corona above the top
of the helmet streamer, and slowly accelerate outward though the LASCO
field of view (2.2-30 $\rsun$). \citet{wang98} suggest that both the
blobs and the plasma sheet represent closed-field material injected
into the slow wind as a result of foot-point exchanges between the
stretched helmet-streamer loops and neighboring open field lines.
According to this model, the ejection of the blobs does not cause any
permanent disruption of the helmet streamer, which remains in a
stretched, quasi-equilibrium state with its cusp at 3 $\rsun$.

\citet{noci97} and \citet{raymond97} observed the equatorial streamer
with UVCS/{\it SOHO} at radii between 1.5 and 5 $\rsun$. Images of the
streamer in O VI $\lambda$1032 show two bright legs separated by a
dark lane. This lane extends radially along the streamer axis up to
3 $\rsun$ \citep[also see][]{strachan2002,frazin2003}. In contrast,
H I Ly$\alpha$ and white-light images do not show such a dark feature.
This led \citet{noci97} to attribute the dark lane to a reduced oxygen
abundance along the streamer axis. \citet{raymond97} further suggested
that this abundance anomaly is due to gravitational settling of oxygen
in the static proton/electron plasma of the closed-field streamer
core. If this interpretation is correct, the closed field lines must
extend up to 3 $\rsun$, consistent with the model of \citet{wang98}.

The O VI images for July 1996 show that the streamer legs converge
towards the equator in the range 2-2.5 $\rsun$ from Sun center. This
suggests that the streamer cusp may be located at about 2 $\rsun$,
somewhat less than implied by the model of \citet{wang98}. We propose
the following scenario. The blobs observed by \citet{sheeley97}
originate at the cusp (2 $\rsun$), but become clearly visible only at
somewhat larger height (3-4 $\rsun$). The plasma within the blobs
originates in the closed-field streamer core, which is affected by
gravitational settling. Therefore, the blobs have low oxygen
abundance compared to the surrounding slow solar wind. As the blobs
move out into the region between 2 and 3 $\rsun$, they contribute to
reduced O VI emission along the axis of the streamer. In this way,
the effects of gravitational settling are transmitted to larger height
via the blobs. In contrast, the open field lines farther away from the
streamer axis always have a slow outflow and therefore are unaffected
by settling.

\citet{strachan2002} measured $\rm O^{5+}$ outflow velocities in
streamers using the Doppler dimming effect \citep[also see][]
{habbal97}. In a latitudinal scan at 2.33 $\rsun$, Strachan
et al.~found no measurable outflow velocity within the streamer ($v <
20$ km/s), but a steep rise in outflow velocity occurs just beyond the
bright streamer legs. The latitudinal width of the streamer, as
defined by the point where the outflow velocity equals 100 km/s, is
about $\pm 20\deg$ \citep[see Fig.~4d of][]{strachan2002}, similar
to the observed width of the slow wind at large distance from the
Sun \citep{Su99}.

\section{Observational Constraints on Global Model}

This section describes the observational constraints on coronal
density, temperature, and magnetic flux to be used in \S 4.

\subsection{Electron Density}

The polarized brightness ($pB$) of the K corona can be used to measure
the coronal electron density \citep{hulst50}.
Here we use the results of \citet{guhatha96} and \citet{sittler99},
who used \emph{Skylab} data from 1973-1974 to derive electron density
profiles for coronal holes and streamers at cycle minimum 
\citep[see also][]{newkirk67,allen73,mj77,saito77}.
Their results are expressed in terms of radial profiles, one for the
polar coronal hole, another for the equatorial streamer. They used
the following expression for the electron density within each
component:
\begin{equation}
N_e(r) = a_1 e^{a_2 /r} r^{-2} [1 + a_3/r + a_4/r^2 + a_5/r^3],
\label{eq:density1}
\end{equation}
where $r$ is the heliocentric distance in units of $\rsun$, and
the parameters $a_1, \cdots , a_5$ are given by \citet{sittler99}.
The density profiles are shown in Figure \ref{fig:dens}. The
\emph{Skylab} data span the height range $1-5\rsun$ and were further
extrapolated using data from the Ulysses mission
\citep{phillips95}. These measurements correspond to select days when
the equatorial streamer belt was seen approximately end-on. This
explains why their streamer densities are somewhat larger than those
of \citet{saito77}, who presented streamer densities averaged over
many days.

\subsection{Electron Temperature}

One method for inferring the electron temperature $T_e$ is to use
spectral line intensity ratios of two lines from the same ion.
\citet{wilhelm98} used data from SUMER/{\it SOHO} to estimate $T_e$ at
radii 1.03-1.6 $\rsun$ in coronal holes at the last cycle minimum
(1996-1997). They use Mg IX 706 {\AA} and 750 {\AA}, and they
conclude that in both the plume and inter-plume regions the electrons
barely reach the canonical temperature of 1 MK. Moreover, $T_e(r)$
falls off rapidly with height. Although their temperature estimates
depend on atomic data that, according to the authors, could be
improved, their work suggests that in polar hole regions the electrons
are  significantly cooler than the ions \citep[see][for a quantitative
analysis on uncertainties of $T_e$ estimates, derived from Be-like
line ratios and using different theoretical methods]{landi01}.

The rates of ionization and recombination of coronal ions decrease
rapidly with distance from the Sun. Therefore, the charge state of
the solar wind is determined in large part by the electron temperature
$T_e$ in the inner corona where the ionization and recombination times
are still short compared to the solar wind expansion time.
{\it In situ} measurements of the solar wind charge state can be
used to estimate the coronal electron temperature. \citet{ko97}
derived polar $T_e$ profiles from observations with
SWICS/Ulysses. Their results can be approximated as a
combination of two power laws \citep[see][]{cranmer99}:
\begin{equation}
T_e (r) = 10^6 \left( 0.35 \ r^{1.1} + 1.9 \ r^{-6.6} \right)^{-1}
~~~ \hbox{[K]}.
\label{eq:Te1}
\end{equation}
\citet{raymond97} and \citet{li98} studied the ionization balance of
various ions in a streamer observed with UVCS/{\it SOHO} in July 1996. 
The results indicate that $T_e$ reaches a maximum value of about 1.6
MK in the streamer core. Due to the high density and low outflow
velocity in streamers \citep[e.g.,][]{frazin2002b}, we expect that
protons and electrons are in thermal equilibrium with each other, so
the electrons may be used as a proxy for the protons.
\citet{raymond97} and \citet{li98} show that their observations are
compatible with hydrostatic equilibrium in the streamer core
\citep[also see]{gibson99}. In this paper we derive the electron
temperature $T_e (r)$ from the observed electron density $N_e (r)$,
assuming hydrostatic equilibrium. 

In this paper we use a generalization of expression (\ref{eq:Te1})
to describe the radial variation of electron temperature at the pole
and the equator:
\begin{equation}
T_e (r) = T_0 \frac{a+1}{a+b r^{\alpha} + (1-b) r^{-\beta}} ,
\label{eq:Te2}
\end{equation}
where $T_0$ is the temperature at the coronal base ($r=1$). The values
of the parameters $a$, $b$, $\alpha$ and $\beta$ are given in Table 1,
and the profiles are shown in Figure \ref{fig:tprof}. The electron
temperature model at the poles (dashed line) has been adjusted to
approximately fit the observational data of \citet{ko97} and
\citet{cranmer99} ({\it triangles}). The electrons are assumed to have
a Maxwellian velocity distribution. The electron temperature at
the equator (thick line) is assumed to be equal to the proton temperature, 
discussed in the next section.

\subsection{Proton Temperature}

Ion temperatures can be estimated by measuring the width of coronal
emission line profiles. Due to rapid charge exchange between protons
and neutral hydrogen, the latter can be used as a proxy for the
protons. \citet{allen98} studied the coupling between neutral hydrogen
and protons by treating the hydrogen atoms as test particles in a
proton-electron background. Their work indicates that the H I velocity
distribution reflects the proton distribution at radii up to about 3
$\rsun$ in the polar regions and even higher in the equatorial
regions. Above 3 $\rsun$ in the polar regions, the H I velocity
distribution ``follows'' the proton distribution and reaches
temperatures about 20\% higher than the protons \citep[see also][]
{olsen94,olsen96}. These results strongly support the use of H I
Ly$\alpha$ profiles to measure the velocity spread of the protons
along the line of sight (LOS). The observed velocity spread includes
both thermal and non-thermal components (such as wave motions), and
may also include a contribution from solar wind expansion in the
direction along the LOS (most emission originates near the point of
closest approach to the Sun, but there is some contribution from
regions behind and in front of the plane of the sky where the solar
wind velocity has a component along the LOS). Therefore, the velocity
spread derived from the observed line width provides only an upper
limit on the proton temperature: $T_p \leq (m_H/k_B) V_{\rm 1/e}^2
/2$, where $V_{\rm 1/e}$ is the observed $1/e$ velocity spread, $m_H$
is the hydrogen mass and $k_B$ is Boltzmann's constant. At the time of
cycle minimum, the polar coronal holes are very large, and polar
observations at $r > 1.5 \rsun$ are unaffected by low-latitude
streamers.

UVCS/{\it SOHO} observations of heavier ions such as $\rm O^{5+}$ show
that the velocity distributions of these ions in coronal holes are
highly {\it anisotropic}: the velocity spread in the radial direction
(as derived from Doppler dimming analysis) is much smaller than that
in the tangential direction \citep{kohl98,cranmer99}. It is unclear
whether this anisotropy also exists for the protons; Doppler dimming
analysis of H I Lyman lines does not provide strong constraints on the
proton parallel velocity in coronal holes. The $\rm O^{5+}$ anisotropy
is believed to be due to the damping of transverse waves (e.g., due to
ion-cyclotron resonance), and this perpendicular heating may also
occur for the protons. In this paper we assume that the protons indeed
have an anisotropic velocity distribution in the low-density polar
regions ($T_{p \perp} > T_{p \parallel}$), but not in the equatorial
region ($T_{p \perp} = T_{p \parallel}$). Since the magnetic field
over the pole is approximately radial and perpendicular to the LOS,
the observed H I Ly$\alpha$ line width provides an estimate for the
proton {\it perpendicular} temperature. Furthermore, we assume that
the proton parallel temperature equals the (isotropic) electron
temperature, $T_{p \parallel} = T_e$, so that the radial variation of
$T_{p \parallel}$ at the pole and equator is given by equation
(\ref{eq:Te2}) with parameter values given in Table 1.  The anisotropy
in proton temperature over the pole is consistent with {\it in-situ}
measurements, which show that $T_p > T_e$ in the solar wind
\citep{marsh82,pilipp87}.

\citet{kohl98} and \citet{cranmer99} analyzed H I Ly$\alpha$ line
profiles observed in polar coronal holes during the past cycle minimum
(1996-1997). Their results indicate proton perpendicular temperatures
up to 6 MK. The observed velocity width ($V_{\rm 1/e}$) increases
rapidly with height from about 190 km/s ($T_{p\perp} \sim 2.2$ MK) at
1.5 $\rsun$, to 240 km/s ($T_{p\perp} \sim 3.5$ MK) at 2.5 $\rsun$,
and then slowly decreases to 250 km/s ($T_{p\perp} \sim 3.8$ MK) at
4 $\rsun$. 
The results are consistent with earlier measurements
from UVCS/{\it Spartan} by \citet{KSG96}, who found peak temperatures
of 5-6 MK at 2.25 $\rsun$ and 3.5 MK at 3.5 $\rsun$ in a polar coronal
hole in 1993. From these several observations, measured temperatures 
at different heights are shown by asterisks in Figure \ref{fig:tprof}.

UVCS/{\it SOHO} observations in the equatorial regions indicate a
roughly constant proton temperature within the core of the streamer
belt. We reanalyzed UVCS data from a super-synoptic campaign during
July 1996 \citep{raymond97}, and find Ly$\alpha$ line widths of order
185 km/s ($T_{p\perp} \sim 2.0$ MK) at 1.5 $\rsun$ and 195 km/s
($T_{p\perp} \sim 2.2$ MK) at 2.6 $\rsun$. At larger heights, the
line widths decrease to 170 km/s ($T_{p\perp} \sim 1.75$ MK) at
3 $\rsun$ and 150 km/s ($T_{p\perp} \sim 1.35$ MK) at 4.5 $\rsun$.
The measurements are shown by the crosses in Figure \ref{fig:tprof}.
Streamer observations by \citet{kohl97}, obtained about a month after
the Raymond observations, exhibit very similar values (2.2 MK at 2
$\rsun$, 1.5 MK at 4 $\rsun$).

In this paper we use the following expression for the proton
perpendicular temperature:
\begin{equation}
T_{p \perp}(r) = T_0 \frac{a+1} {a+b r^{\alpha} + (1-b) r^{-\beta}} +
T_1 \frac{(r-1)^2 ~ e^{-(r-1)/\Delta r}}
{(r_{max}-1)^2 ~ e^{-(r_{max}-1)/\Delta r}} ,
\label{eq:Tperp}
\end{equation}
The first term is similar to that used for the electron temperature,
but the second term allows us to impose a further increase in
temperature over a limited range of heights (of order $\Delta r$),
consistent with UVCS observations. The values of the parameters
at the pole and at the equator are given in Table 1.

In summary, our temperature models present the following main
features: (1) at the pole, strong anisotropy for protons, and
electron temperature much lower than $T_{p \perp}$; (2) at the
equator, hydrostatic equilibrium isotropic velocity distributions and
thermal equilibrium between species. Note that $T_0$ and $\beta$
are the same for all temperature models, so that $T_{p \perp} =
T_{p \parallel} = T_e$ at low heights in the corona.

\subsection{Magnetic Flux at Coronal Base}

Figure \ref{fig:magn} shows the longitude-averaged radial magnetic
field $B_r$ as function of latitude at the solar surface ($r = 1$).
These data are derived from the Kitt Peak synoptic map for July 1996,
near the time of cycle minimum. There are no large active regions on
the Sun at this time. The {\it full} curve in Fig.~\ref{fig:magn}
is a fit of the form
\begin{equation}
B_r (\theta) = B_0 \cos^p \theta ,
\label{eq:Bbase}
\end{equation}
where $p=7$ and $B_0$ = +10 G. Note that $B_r$ decreases monotonically
from about +10 G in the North to -10 G in the South. At latitudes
between $\pm 50$ degrees, the average radial field is very small
(less than 2 G). Therefore, the large-scale coronal field is
dominated by the polar fields.

\section{Coronal Model with Anisotropic Gas Pressure}

At cycle minimum, the corona presents a relatively ordered structure
with high latitude coronal holes and an equatorial streamer belt.
The corona is essentially axisymmetric and stable for many months
\citep[e.g.,][]{gibson01}. In the following we describe an
axisymmetric model of the corona; an earlier version of this
model was described by \citet{Vas99a} and \citet{Vas99b}. We use a
spherical coordinate system $(r,\theta,\phi)$ and assume that all
scalar quantities are independent of azimuth $\phi$. The
magnetic field $\bf B$ is assumed to lie in the meridional plane:
\begin{equation}
{\bf B} = \nabla \times {\bf A} = \frac{1}{r \sin \theta} \left(
\frac{1}{r} \frac{\partial A}{\partial \theta} \hat{\bf e}_r -
\frac{\partial A}{\partial r} \hat{\bf e}_\theta \right) ,
\label{eq:Bvec}
\end{equation}
where ${\bf A} = A_\phi \hat{\bf e}_\phi$ is the vector potential,
and $A = A_\phi r \sin \theta$ is the field-line variable
(note that $A$ = constant along field lines). This variable increases
from $A = 0$ along the polar axis to $A = A_{eq}$ at the equator ($r =
\rsun$, $\theta = \pi / 2$). There is a critical field line $A_c$ that
forms the boundary between open and closed magnetic regions; the
field lines with $0 < A < A_c$ are open, while those with $A_c <
A < A_{eq}$ are closed. Note that $A_c / A_{eq}$ is the fraction of
surface flux that is open. In this paper we assume $A_c/A_{eq}$ =
0.80. The magnetic configuration is illustrated in Figure
\ref{fig:sketch}.

Rather than solving an energy equation for the coronal plasma, our
semi-empirical model is based on observed temperature distributions.
The electron and proton temperatures ($T_e$, $T_{p \parallel}$ and
$T_{p \perp}$) are interpolated between the pole and equator:
\begin{equation}
T (r,A) = T_{pole}(r) + [ T_{equa}(r) - T_{pole}(r) ] \Phi (A) ,
\label{eq:temp}
\end{equation}
where $T_{pole}(r)$ and $T_{equa}(r)$ are the observed profiles
described in \S 3, and $\Phi(A)$ is a function of the field-line
variable that increases monotonically from $\Phi(0) \approx 0$ at the
pole to $\Phi(A_c) \approx 1$ at the equator. We use the following
expression for $\Phi(A)$:
\begin{equation}
\Phi(A)= \frac{1}{1+\exp{[-(A-A_h)/A_w]}} ,
\label{eq:phiA}
\end{equation}
where $A_h$ is the mid-point of the transition [$\Phi(A_h) =
\onehalf$], and $A_w$ measures the width of the transition. In this
paper we use $A_h/A_c$ = 0.70 and $A_w / (A_c-A_h)$ = 0.15. The
function $\Phi(A)$ is shown in Figure \ref{fig:phiA}.

The coronal magnetic and velocity fields are computed by solving the
momentum equation:
\begin{equation}
\nabla \cdot \left( \bar{\bf P} + \rho {\bf v v} \right) =
- \rho \nabla \Phi_g + \frac{1}{4\pi} (\nabla \times {\bf B})
\times {\bf B},
\label{eq:mom}
\end{equation}
where $\bar{\bf P}$ is the pressure tensor, $\rho$ is the mass
density, $\bf v$ is the outflow velocity, $\Phi_g (r) = -G\MS /r$ is
the gravitational potential. The steady-flow condition requires that
$\bf v \parallel \bf B$, and conservation of mass requires that
$\rho v / B$ is constant along field lines. The pressure tensor is
anisotropic:
\begin{equation}
\bar{\bf P} = \ppar \shat \shat + \pper \left( \bar{\bf I} -
\shat \shat \right) ,
\label{eq:pressure}
\end{equation}
where $\bar{\bf I}$ is the unit tensor, $\shat$ is the unit vector
along ${\bf B}$, and $\ppar$ and $\pper$ are the parallel and
perpendicular pressures:
\begin{eqnarray}
\ppar & = & N_e k_B ( T_e + T_{p \parallel} ) , \\
\pper & = & N_e k_B ( T_e + T_{p \perp} ) .
\end{eqnarray}
Here we assume a pure hydrogen plasma. The left-hand side of equation
(\ref{eq:mom}) can be written as 
\begin{equation}
\nabla \cdot \left( \bar{\bf P} + \rho {\bf v v} \right) =
\nabla \pper + B \frac{d}{ds} \left( \frac{\delta p}{B} \right) \shat
+ \delta p \frac{d \shat}{ds} ,
\label{eq:part1}
\end{equation}
where $s$ measures distance along the field lines, $B \equiv
| {\bf B} |$, $\delta p \equiv \ppar - \pper + \rho v^2$, and we
used $\nabla \cdot {\bf B} = 0$. The component of the momentum
equation (\ref{eq:mom}) parallel to $\shat$ reads
\begin{equation}
\frac{d \pper}{ds} + B \frac{d}{ds} \left( \frac{\delta p}{B} \right)
= - \rho \frac{d \Phi_g}{ds} ,
\label{eq:mompara}
\end{equation}
and the solution of this equation will be further discussed in \S 4.2.
We now introduce the function $\pbig (r,A)$ describing the dependence
of $\pper$ on the field-line variable: $\pper (r,\theta) = \pbig
[r,A(r,\theta)]$. Then the gradient of $\pper$ is
\begin{equation}
\nabla \pper = \frac{\partial \pbig}{\partial r} \hat{\bf e}_r +
\frac{\partial \pbig}{\partial A} \nabla A ,
\label{eq:pgrad}
\end{equation}
where the partial derivatives on the RHS are taken at constant $A$ and
constant $r$, respectively. It follows that $d \pper /ds =  (\partial
\pbig / \partial r) \cos \gamma$, where $\gamma$ is the angle between
$\shat$ and $\hat{\bf e}_r$, and similarly $d \Phi_g /ds = 
(d \Phi_g /dr) \cos \gamma$. Inserting these expressions into equation
(\ref{eq:mompara}), we obtain
\begin{equation}
\frac{\partial \pbig}{\partial r} + \rho \frac{d \Phi_g}{dr} =
- \frac{B}{\cos \gamma} \frac{d}{ds} \left( \frac{\delta p}{B}
\right) ,
\label{eq:part2}
\end{equation}
and combining equation (\ref{eq:mom}), (\ref{eq:part1}),
(\ref{eq:pgrad}) and (\ref{eq:part2}) yields
\begin{equation}
\frac{\partial \pbig}{\partial A} \nabla A + \frac{B} {\cos \gamma}
\frac{d}{ds} \left( \frac{\delta p}{B} \right) \left( \shat \cos
\gamma  - \hat{\bf e}_r \right) + \delta p \frac{d \shat}{ds} =
\frac{( \nabla \times {\bf B} )_\phi}{4\pi} (- B_\theta \hat{\bf e}_r
+ B_r \hat{\bf e}_\theta ) .
\label{eq:momperp1}
\end{equation}
In streamers the pressure anisotropy and outflow velocity are small,
so the terms involving $\delta p$ can be neglected. The same is true
in coronal holes because in these regions the magnetic field is nearly
radial and the vectors $(\shat \cos \gamma  - \hat{\bf e}_r)$ and
$d \shat /ds$ vanish for radial fields. Therefore, $\delta p$ plays
only a minor role in the perpendicular force balance, and we will
neglect its effects in equation (\ref{eq:momperp1}). Then the
perpendicular force balance (\ref{eq:momperp1}) reduces to
\begin{equation}
\frac{( \nabla \times {\bf B} )_\phi}{4\pi} = r \sin \theta
\frac{\partial \pbig}{\partial A} ,
\label{eq:momperp2}
\end{equation}
where we used equation (\ref{eq:Bvec}). The solution of equation
(\ref{eq:momperp2}) will be discussed in \S 4.1.

We use an iterative method for solving equations (\ref{eq:part2})
and (\ref{eq:momperp2}). The method can be summarized as follows.
We first construct a preliminary pressure model $\pbig (r,A)$ by
interpolating the observed perpendicular pressure between pole and
equator, using an expression similar to equation (\ref{eq:temp}).
Then we compute the field-line variable $A(r,\theta)$ by solving the
perpendicular force balance as described below in \S 4.1; this gives
us a first guess for the shape of the magnetic field lines. Then we
solve the parallel force balance separately along many field lines
as described in \S 4.2. This yields the mass density $\rho(s)$ and
velocity $v(s)$ along each field line, which are then remapped to
produce the density $\rho(r,\theta)$ and velocity $v(r,\theta)$.
We also update the perpendicular pressure $\pbig (r,A)$, which is then
used in the next iteration to recompute the field-line variable
$A(r,\theta)$. This process is repeated until convergence is
achieved. Note that in the first iteration $\pbig (r,A)$ is specified
analytically, whereas in subsequent iterations $\pbig (r,A)$ is
obtained by numerical interpolation.

\subsection{Perpendicular force balance}

To solve equation (\ref{eq:momperp2}) for the perpendicular force
balance, we introduce the following integral over the coronal volume:
\begin{equation}
L = \int_0^{\pi/2} \int_{\rsun}^{\rmax} \left[ \frac{B^2}{8\pi} -
\pbig (r,A) \right] r^2 \sin \theta \ dr \ d\theta ,
\label{eq:Lagr}
\end{equation}
where only one hemisphere of the Sun is considered, and $\rmax$
= 10 $\rsun$ is the outer radius of the computational domain.
The magnetic field $\bf B$ is related to $A_\phi$ via equation
(\ref{eq:Bvec}), and $A \equiv A_\phi r \sin \theta$. At each step in
our iterative procedure, $\pbig (r,A)$ is a known function, so $L$
depends only on the unknown function $A_\phi (r,\theta)$, or
equivalently, $A(r,\theta)$. We assume the following boundary
conditions for $A(r,\theta)$:
\begin{eqnarray}
A(\rsun,\theta) & = & \frac{B_0 \rsun^2}{p+1} \left( 1 - \cos^{p+1}
\theta \right) , \label{eq:Abase} \\
\frac{\partial A}{\partial r} (R_{max},\theta) & = & 0 , \\
A(r,0) & = & 0, \\
A(r,\pi/2) & = & A_c ~~~ \hbox{for} ~ r > \rcusp , \label{eq:Aeq1} \\
\frac{\partial A}{\partial \theta} (r,\pi/2) & = & 0
~~~ \hbox{for} ~ r < \rcusp . \label{eq:Aeq2} 
\end{eqnarray}
Equation (\ref{eq:Abase}) is derived from the surface flux
distribution, equation (\ref{eq:Bbase}), and the magnetic field is
assumed to be radial at the outer boundary, $r = \rmax$. Equations
(\ref{eq:Aeq1}) and (\ref{eq:Aeq2}) describe the boundary conditions
at the equator. For $r > \rcusp$, there is a current sheet at the
equator separating the open fields from the two hemispheres, hence the
magnetic field just above the equator is radial. For $r < \rcusp$ the
magnetic field is closed over the equator, so the field is
perpendicular to the equatorial plane.

We now show that the function $A_\phi (r,\theta)$ for which $L$
reaches its {\it minimum} value is a solution of equation
(\ref{eq:momperp2}). Let $\delta A_\phi (r,\theta)$ be an arbitrary
variation of $A_\phi (r,\theta)$, then the corresponding change in
$L$ is
\begin{equation}
\delta L = \int_0^{\pi/2} \int_{\rsun}^{\rmax} \left[
\frac{( \nabla \times {\bf B} )_\phi} {4\pi} - r \sin \theta
\frac{\partial \pbig} {\partial A} \right] \delta A_\phi (r,\theta)
~ r^2 \sin \theta \ dr \  d\theta ,
\label{eq:delta_Lagr}
\end{equation}
where the boundary conditions (\ref{eq:Abase}) - (\ref{eq:Aeq2}) have
been used. At the minimum, $\delta L$ vanishes for any function
$\delta A_\phi (r,\theta)$, so the quantity in square brackets in
equation (\ref{eq:delta_Lagr}) must vanish. This is precisely the
condition for perpendicular force balance, equation
(\ref{eq:momperp2}).

The function $A_\phi (r,\theta)$ is discretized on a non-uniform grid
in $r$ and $\theta$ with 180 points in each direction. The
minimization of $L$ is performed using the Polak-Ribiere version of
the conjugate gradient method \citep{NR}. This is an iterative
method for adjusting the values of $A_\phi$ at the $180^2$ grid
points until the minimum of $L$ is reached. The cusp radius $\rcusp$ is
also allowed to change in order to obtain the lowest possible $L$,
but the amount of open magnetic flux ($A_c/A_{eq}$) is held fixed.

\subsection{Parallel force balance}

The parallel component of the momentum equation (\ref{eq:mompara}) can
be written in the following form:
\begin{equation}
B \frac{d}{ds} \left[ \frac{\rho}{B} \left( \cpar^2 + v^2 \right)
\right] + \rho \cper^2 \frac{1}{B} \frac{dB}{ds} = - \rho
\frac{d \Phi_g}{ds} ,
\label{eq:wind1}
\end{equation}
where $c_\parallel$ and $c_\perp$ are the parallel and perpendicular
sound speeds:
\begin{equation}
c_\parper^2 \equiv \frac{p_\parper}{\rho} =
\frac{k_B (T_{p~\parper} + T_e)} {m_p} .
\end{equation}
Using $\rho v/B$ = constant along field lines, we can eliminate the
density from equation (\ref{eq:wind1}):
\begin{equation}
\left( 1 - \frac{\cpar^2}{v^2} \right) v \frac{dv}{ds} =
\frac{df}{ds} , 
\label{eq:wind2}
\end{equation}
where
\begin{equation}
f(s) \equiv \frac{G\MS}{r(s)} - \cpar^2(s) -
\int_0^s \frac{\cper^2(s')}{B(s')} \frac{dB}{ds'} ~ ds' .
\label{eq:fs}
\end{equation}
At each step of our iterative procedure we have estimates for $B(s)$,
$r(s)$, $T_{p \parallel}(s)$, $T_{p \perp}(s)$ and $T_e(s)$ on any open
field line, which allows us to compute the function $f(s)$. According
to equation (\ref{eq:wind2}), the sonic point ($v = \cpar$) is located
at an extremum of $f(s)$. The velocity $v(s)$ is found by inward and
outward integration of equation (\ref{eq:wind2}), starting at the
sonic point. In practice we find that, in order to obtain a valid
solution over the entire height range ($\rsun < r < \rmax$), the sonic
point must be located at the {\it global minimum} of $f(s)$. The
density $\rho(s)$ is computed from mass flux conservation and the
boundary condition for the density at the coronal base (see \S 3).
This process is repeated for 180 different open field lines, and the
results are remapped to obtain the density $\rho(r,\theta)$ and
velocity $v(r,\theta)$ on the $(r,\theta)$ grid.

\section{Results}

We iterated the parallel and perpendicular force balance equations as
described in \S 4. The process was repeated until the maximum change
in $\pbig (r,A)$ between iterations is less than $1\%$; this required
20 iterations. The results for the first and last iterations are shown
in Figures \ref{fig:fld}a and \ref{fig:fld}b, respectively.
The {\it thin} curves represent magnetic field lines, and the
{\it thick} curves are contours of $\beta(r,\theta)$, the ratio of
perpendicular gas pressure $p_\perp$ and magnetic pressure $B^2/8\pi$.
The cusp height in the final solution is $\rcusp \approx 1.95 \rsun$.
Note that the magnetic structure changes little between the first and
last iterations.

Our model yields $\beta \ll 1$ in the polar regions, but $\beta > 1$
in the equatorial streamer, especially near the streamer cusp. 
This supports the idea that the streamer is magnetically contained by
the strong polar fields that surround it on either side
\citep{SuGaNe99}. We find that $\beta > 1$ throughout the closed-field
region of the streamer. However, the variation with height is not
monotonic: $\beta$ has peaks at both the cusp and the streamer base,
and lower values at intermediate heights. Our results are similar to
those of \citet{li98}, who estimated $\beta$ based on UVCS/{\it SOHO}
and SXT/{\it Yohkoh} observations of streamers (July 1996) in
combination with potential field extrapolation of the photospheric
magnetic field.  Their estimates indicate $\beta \sim 5$ at 1.15
$\rsun$ and $\beta \sim 3$ at 1.50 $\rsun$, similar to the values
found here. High values of $\beta$ were also found in MHD models that
include heat and momentum deposition in the corona
\citep{WWSP98,Su96}.

Figure \ref{fig:velo} shows the outflow velocity for the final model.
The bright region is the fast solar wind emanating from the coronal
hole, and the dark region is the slow wind that flows along the open
field lines within the streamer. The maximum velocities are about
450 km/s for the fast wind and 190 km/s for the slow wind. This is
somewhat smaller than the observed {\it in-situ} values at 1 AU
(800 and 400 km/s, respectively). We attribute this difference to
the fact that our model does not include any momentum deposition
effects other than the mirror force. However, the ratio of fast and
slow wind speeds is about 2.5 in our model, consistent with {\it
in-situ} measurements.

Figure \ref{fig:across} shows three quantities measured across
magnetic field lines: the temperature anisotropy at 3 $\rsun$,
the radial position $r_s$ of the sonic point, and the asymptotic wind
speed. Note that the variation of wind speed closely follows that of
the temperature anisotropy (compare top and bottom panels). This
indicates that the fast wind is driven by the high perpendicular
temperatures in the coronal hole, and the decrease in wind speed at
the edge of the hole is mainly due to the decrease in temperature
anisotropy.  The middle panel shows that there is a sudden jump in
sonic-point radius once the transition from high to low temperature
anisotropy is nearly complete. The jump occurs at $A/A_c = 0.77$ and
is due to the appearance of a second (lower) minimum in the function
$f(s)$ at a height well beyond the streamer cusp (see below). As a
result, the sonic-point radius changes discontinuously from about 1.5
$\rsun$ in the coronal hole to about 5.5 $\rsun$ in the streamer. Such
large values for the sonic-point height in the streamer were found
earlier by \citet{wang94} and \citet{chen2001}. Most of the decrease
in wind speed occurs for $A/A_c$ between 0.6 and 0.77, well before the
jump in sonic-point height, and the asymptotic wind speed does not
change significantly at the jump.

To show the transition between fast and slow wind more clearly,
Figure \ref{fig:trans} shows various quantities along field lines
near the fast-slow boundary (the same line styles are used in
all panels). Panel (a) show the shapes of the field lines in the
meridional plane. Panel (b) shows the function $f(s)$ defined in
equation (\ref{eq:fs}) and plotted as function of radial distance
$r$ for four field lines. Note that the {\it triple-dot-dashed}
curve has a minimum at about 1.5 $\rsun$, whereas the {\it full}
curve has two minima, one at 1.7 $\rsun$ and another at 5.5 $\rsun$.
As shown in \S 4.2, these minima indicate possible positions of the
sonic point ($v = c_\parallel$). Panel (d) shows the radial
positions $r_{\rm min}$ of such minima as function of field-line
variable $A$. In general, a global solution of the wind equation
(\ref{eq:wind2}) can be found only when the integration is started
at the {\it global} minimum, i.e., the minimum with the lowest value
of $f(r_{\rm min})$. Therefore, as we move from the
{\it triple-dot-dashed} field lines to the {\it dot-dashed}
line, there is a discontinuous jump in the height of the sonic point.

Figure~\ref{fig:trans}c shows the outflow velocity $v(r)$ along field
lines near the fast-slow boundary. The {\it dotted} curve in this
panel is the wind solution along the last open field line, $A = A_c$.
As we move from the coronal hole into the streamer, the wind speed at
large height closely follows the decrease in perpendicular
temperature, which mainly occurs between the two {\it full-thick}
profiles. However, a further decrease occurs at lower heights
($r \sim 2$ $\rsun$) between the {\it triple-dot-dashed} and
{\it dot-dashed} field lines due to the jump in height of the sonic
point. We conclude that, unlike the terminal speed, the low-height
behavior of $v(r)$ is strongly affected by the height of the sonic
point. Such a region of stagnated flow was found earlier by many
authors \citep[e.g.,][]{wang94,SuNe99,SuNe02,chen2001}.

Figure~\ref{fig:trans}e shows the non-radial expansion factor of the
slow-wind flow tubes, $f_{\rm exp}(r) = (\rsun/r)^2 B(\rsun)/B(r) $,
where $B(\rsun)$ in the field strength at the coronal base for each
field line. Note that $f_{\rm exp}(r)$ is a non-monotonic function of
radius for these field lines. The peak in $f_{\rm exp}(r)$ is due to
the fact that these field lines pass close to the streamer cusp where
$B \approx 0$. In contrast, $f_{\rm exp}(r)$ increases monotonically
for field lines in the coronal hole (not shown). It is remarkable to
see that this ``cusp effect'' occurs at large distance from the cusp:
the peak in $f_{\rm exp}(r)$ occurs for all field lines within about
1 $\rsun$ from the equatorial plane (see Figure~\ref{fig:trans}a).
The {\it triple-dot-dashed} field line at which the peak in $f_{\rm
exp}(r)$ first develops is close to the field line where the jump in
sonic-point height occurs. This is consistent with the suggestion by
\citet{SuNe99,SuNe02}, \citet{chen2002} and others that the existence
of the slow wind is due to the non-monotonic behavior of
$f_{\rm exp} (r)$. We suggest here that the decrease in perpendicular
temperature as we approach the equatorial plane also plays an
important role.

To determine whether the ``cusp effect'' is due to the effects of gas
pressure on the expansion of the flow tubes, we computed a partially
open potential magnetic field with the same photospheric boundary
conditions [see equation (\ref{eq:Bbase})] and a thin current sheet
in the equatorial plane for $r > 1.95 \rsun$. The calculation is based
on the work of \citet{low86}. First, the surface flux distribution
is expressed in terms of four Legendre polynomials $P_n(x)$ with $n$
= 1, 3, 5 and 7. Then the field is extrapolated into the corona using
equations (B3) and (B4) of \citet{low86} for $n$ = 1 and 3 (for $n$ =
5 and 7 the effect of the current sheet can be neglected). We found
that $f_{\rm exp}(r)$ has a peak for all open field lines that pass
below the point $(x,z)$ = (3.0,1.4) $\rsun$, where $x$ is the distance
from the rotation axis, and $z$ is the height above the equatorial
plane.  Therefore, the effect of the streamer cusp -located at
the point $(x,z)$ = (1.95,0) $\rsun$- is present even in a potential
field, and its spatial extent is similar to what was found in the
numerical model (i.e., the effect is not due to the finite plasma
$\beta$).

For the value of $A_h$ chosen in this paper ($A_h/A_c$ = 0.7), the
transition between fast and slow wind far from the Sun occurs at
latitudes of $\pm 20\deg$. This is consistent with Ulysses
observations taken at the time of cycle minimum \citep{Su99}. At lower
heights, the low speeds in the streamer legs found in our model are
consistent with UVCS measurements of $\rm O^{5+}$ outflow velocity
using the Doppler dimming technique \citep{strachan2002}. In a
latitudinal scan at 2.33 $\rsun$, \citet{strachan2002} found no
measurable outflow velocity in the streamer ($v < 20$ km/s), but a
steep rise in velocity occurs just beyond the bright streamer
legs. The observed latitudinal half-width of the streamer at 2.33
$\rsun$, as defined by the points where the outflow velocity equals
100 km/s, is about $20\deg$. Our model predicts a half-width of
$22\deg$. Therefore, the predicted width of the streamer is roughly
consistent with observations at both large and small heights.

It should be noted that we use a realistic photospheric flux
distribution that is peaked at the pole, $B_r (\rsun,\theta) = 10
\cos^7 \theta$ [G], whereas \citet{SuGaNe99} and \citet{chen2001}
assume a bipolar distribution, $B_r (\rsun, \theta) \propto \cos
\theta$. For a given fraction of open magnetic flux, the assumed
flux distribution has a significant effect on the size of the polar
coronal hole, the latitudinal width of the streamer core, and the
magnitude of the non-radial expansion factor. However, potential field
models indicate that the spatial extent of the cusp effect is similar
in the two cases.

\section{Comparison with Visible Light and Ly$\alpha$ Observations}

The coronal density model can be used to predict the polarization
brightness ($pB$) of the visible light that is scattered by free
electrons in the corona (Thomson scattering). The $pB$ is an
integral along the LOS \citep{hulst50}:
\begin{equation}
pB(r_0) = \frac{3}{16} \sigma_T \overline{B}_\odot 
\int_{\rm LOS} N_e(x) \left( \frac{r_0}{r} \right)^2
\frac{(1-u') \tilde{A}(r) + u' \tilde{B}(r)} {1-u'/3} ~ dx ,
\label{eq:pB}
\end{equation}
where $x$ is the distance along the LOS, $r_0$ is the projected
radial distance from Sun center, $r = \sqrt{r_0^2+x^2}$ is the true
radial distance, $\sigma_T = 6.65 \times 10^{-25}$ $\rm cm^2$ is
the Thomson scattering cross section, $\overline{B}_\odot = 1.97
\times 10^{10}$ $\rm erg~ cm^{-2} s^{-1} sr^{-1}$ is the mean 
disk intensity, and $u' = 0.63$ is the limb darkening coefficient.
The functions $\tilde{A}(r)$ and $\tilde{B}(r)$ are given by
\citet{hulst50}. We used our axisymmetric model of the electron
density distribution to compute synthetic $pB$ images for various
angles $\alpha$ between the LOS and the solar equatorial plane.
Figure \ref{fig:pB1} shows the results for $\alpha = 0$ and $\alpha
= 30\deg$; the latter represents a view of the Sun from above
the ecliptic plane. We have over-plotted selected field lines in
the plane of the sky ({\it white} curves) and contours of the
ratio of $pB$ and its maximum value at the equatorial base ({\it
black} curves). The equatorial streamer clearly stands out in the
image with $\alpha = 0$, but is less distinct for $\alpha = 30\deg$.

To compare these results with observations, Figure~\ref{fig:pB2}a
shows the radial $pB$ profiles along the pole and equator, and
Figure~\ref{fig:pB2}b shows latitudinal profiles at three different
heights ($r_0$ = 1.15, 1.5 and 2.5 $\rsun$). The {\it full} curves
correspond to the edge-on view ($\alpha = 0$). The symbols represent
$pB$ measurements obtained with the Mauna Loa Mark 3 coronagraph and
LASCO/{\it SOHO} \citep{gibson99, guhatha99} and with the
visible-light channel on UVCS/{\it SOHO} \citep{cranmer99}. The data
were obtained in 1996 July and August when the equatorial streamer
belt is seen approximately edge-on \citep{raymond97}. Our model fits
the observations quite well, although the predicted latitudinal width
of the streamer at 2.5 $\rsun$ is somewhat larger than observed.  The
{\it dashed} curve in Figure~\ref{fig:pB2}a shows the radial profile
along the pole for the out-of-ecliptic view ($\alpha = 30\deg$). Note
that there is a large increase in $pB$ compared to the edge-on
view. This is due to the fact that the LOS crosses the enhanced
density equatorial streamer, and is not representative of the
available observations of the polar $pB$ profile.

We also computed the intensity of the strongest UV coronal emission
line, H I 1216 {\AA} (Ly$\alpha$), which is formed almost entirely
by resonant scattering of chromospheric Ly$\alpha$ radiation by
neutral hydrogen atoms in the corona \citep{noci87,cranmer99}.
The integrated intensity of the scattered Ly$\alpha$ is
\begin{equation}
I = \frac{h\nu_0}{4\pi} B_{12} \int_{\rm LOS} N_H(x) \Omega(r)
\int I_\odot (\nu + \delta\nu) \phi(\nu-\nu_0) ~ d\nu ~ dx ,
\label{eq:Lya}
\end{equation}
where $x$ measures distance along the LOS, $N_H(x)$ is the neutral
hydrogen density, $r$ is the radial distance from Sun center at any
point along the LOS, $\Omega (r) \equiv [1-\sqrt{1-(\rsun/r)^2}]/2$
is a geometric dilution factor, $I_\odot (\nu)$ is the chromospheric
intensity (in $\rm erg ~ cm^{-2} s^{-1} Hz^{-1} sr^{-1}$) as function
of frequency $\nu$ (in Hz), $\nu_0 = 2.47 \times 10^{15}$ Hz is the
line center frequency, $\dn \equiv \nu_0 v_r/c$ is the Doppler shift,
$v_r$ is the radial component of the outflow velocity,
$\phi(\nu-\nu_0)$ is the scattering profile of the coronal atoms (due
to their thermal and non-thermal velocities), and $B_{12}$ is the
Einstein coefficient.  For a derivation of this expression and
discussion of the relevant approximations, see \citet{noci87}.
The neutral hydrogen density $N_H$ is computed using the collisional
ionization rates of \citet{SW91} and the recombination rates of
\citet{hummer94}.

We computed Ly$\alpha$ images for two different viewing angles, using
densities, temperatures and velocities from the coronal model. The
results are shown in Figure~\ref{fig:Lya1}. Although these images
look similar to those for polarization brightness, the Ly$\alpha$
intensity is somewhat sensitive to the outflow velocity. This is
shown more clearly in Figure~\ref{fig:Lya2} where we plot the radial
profiles of Ly$\alpha$ intensity with and without the Doppler dimming
effect ({\it full} and {\it dashed} curves, respectively). Note that
there is a significant difference between predicted intensities with
and without dimming for the polar region, but not for the equatorial
region. Therefore, unlike $pB$ measurements Ly$\alpha$ intensities
provide constraints on the outflow velocity in the acceleration region
of the solar wind.

Figure \ref{fig:Lya2} also compares our model predictions with
measurements from UVCS/{\it SOHO}. The {\it diamonds} represent
measurements taken along the streamer axis, based on our reanalysis of
data obtained on 1996 July 26 \citep{raymond97}. The {\it triangles}
are measurements along the
polar axis based on observations from November 1996 to April 1997
\citep{cranmer99, danuta99}. Note that the predicted Ly$\alpha$
intensities fit the observations quite well. This suggests that the
model gives a reasonable representation of the velocity field in the
inner part of the corona where the main acceleration of the solar wind
takes place.

\section{Discussion}

We developed a stationary, axisymmetric MHD model for the global
corona. The model includes a description of the temperature, density,
velocity and magnetic field as function of latitude and radius up to
10 $\rsun$ from sun center. The velocity and magnetic fields are
obtained by solving the parallel and perpendicular force balance
equations, including the effects of inertia, anisotropic gas pressure,
gravity and Lorentz forces. The temperature models are based on
observational data from UVCS/{\it SOHO}, and the magnetic flux
distribution at the coronal base is taken from NSO/Kitt Peak synoptic
maps. The model reproduces the main features of the global corona
at the time of cycle minimum. We find that the high perpendicular
temperature of the protons in the coronal hole plays a major role in
driving the fast solar wind. In the streamer we find low outflow
velocity and high plasma $\beta$, consistent with earlier results
\citep{Su96,WWSP98,li98}.

In our model the wind equation (\ref{eq:wind2}) is solved separately
for many field lines. The sonic point along each field line occurs
at a minimum of the function $f(s)$ that appears on the RHS of the
wind equation [see equation (\ref{eq:fs})]. The transition from fast
to slow wind occurs at an open field line characterized by $A_h =
0.7 A_c$, where $A_c$ is the boundary between open and closed fields.
The value of $A_h$ was adjusted to obtain the correct latitudinal
width of the slow-wind region both at large distance from the Sun
\citep{Su99} and in the corona at 2.33 $\rsun$ \citep{strachan2002}.
In our model this transition is associated with two effects:
(1) the decrease of proton perpendicular temperature $\Tpper (r,A)$
as we approach the equatorial plane, and (2) the appearance of a
peak in the non-radial expansion factor $f_{\rm exp}(r)$ for field
lines that pass close to the streamer cusp (``cusp effect'').
The cusp effect is present at surprisingly large distances from the
cusp ($\sim 1$ $\rsun$) and is present even in potential-field models,
so it is not a consequence of finite plasma $\beta$.
The combination of decreasing temperature anisotropy and cusp effect
causes the global minimum of $f(s)$ (and therefore the sonic point)
to occur well beyond the cusp, and produces low outflow velocity near
the cusp. These results are consistent with the suggestion by
\citet{noci97} that the slow wind is due to special properties of
the geometric spreading along the open field lines that pass near
the streamer core \citep[also see][]{wang90,wang94,wang97b,chen2001,
chen2002,SuNe99,SuNe02}.

Our model does not provide a physical explanation for the temperature
decrease at the fast-slow boundary, and therefore cannot explain 
{\it why} the boundary occurs at $A_h = 0.7 A_c$. Understanding
the physics of the fast-slow transition will require more detailed
analysis of the energy balance of the coronal plasma, including the
physical processes by which the temperature anisotropy of the protons
is maintained. We speculate that proton
perpendicular heating (by dissipation of transverse MHD waves) occurs
in both the fast and slow winds, perhaps at roughly equal rates.
However, the resulting temperature anisotropy $\Tpper / \Tppar$ may
be quite different in the two cases. In the fast wind, proton-proton
collisions are less frequent due to the lower density, so the
deviations from Maxwellian velocity distributions are larger than
in the slow wind. Clearly, to understand why the fast-slow transition
occurs at $A_h = 0.7 A_c$ will require multi-dimensional models of
wave heating and energy balance such a those developed by
\citet{chen2001}.

\acknowledgments 

We express our thankfulness to the anonymous referee for his/her valuable 
comments on the paper, which helped to clarify the manuscript. We also
thank John Mariska for his useful suggestions on the manuscript.
This work was funded by the Smithsonian Astrophysical Observatory,
and by the CONICET (The Argentinean National Council for Scientific
and Technological Research). We also thank \emph{Fundaci\'on Antorchas}
(from Argentina) for partial support through grant 14056-20.

%
%

\clearpage

\figcaption[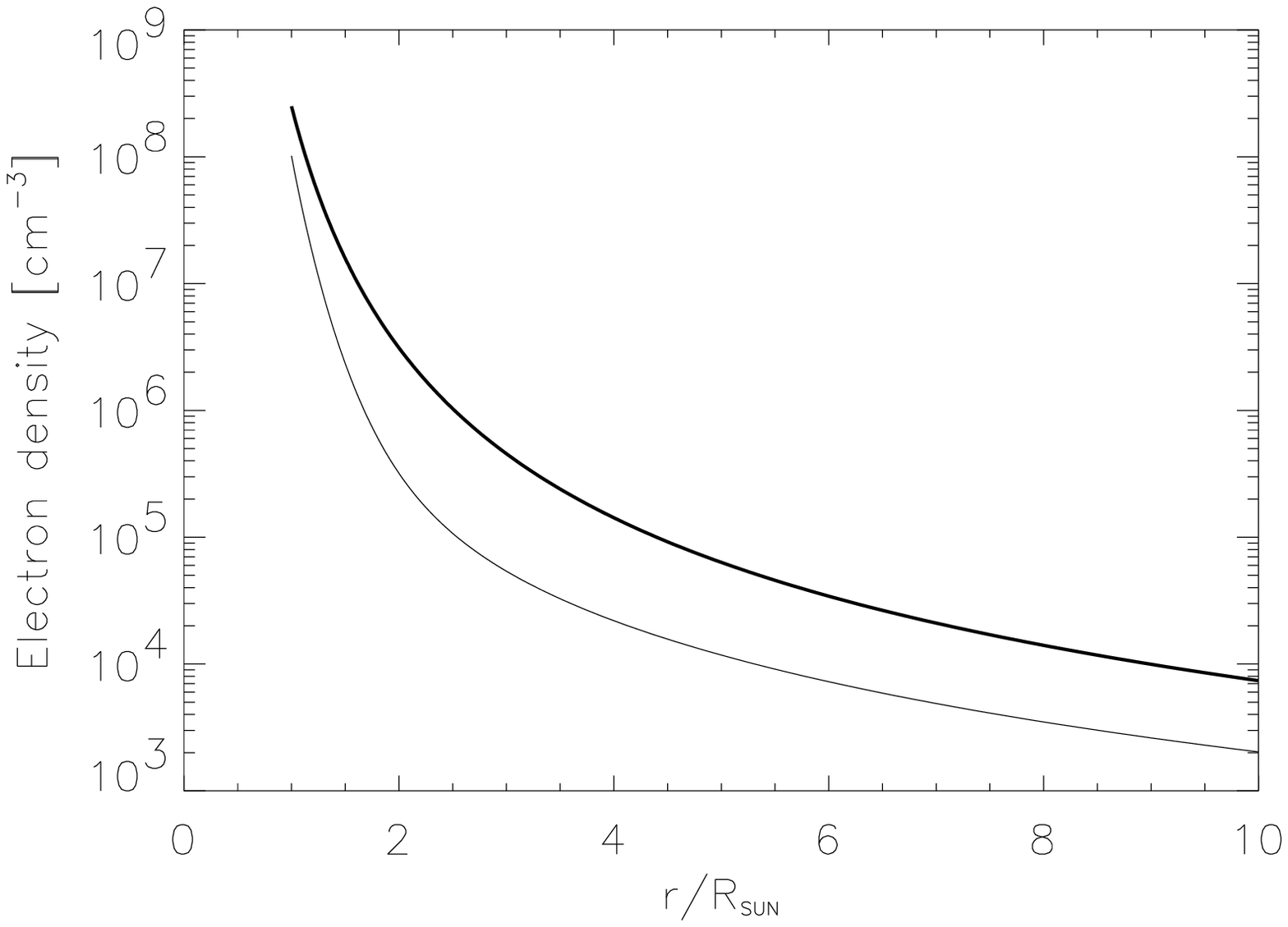]
{Electron density profiles derived from \emph{Skylab} observations
for the pole ({\it thin} curve) and the equator ({\it thick} curve).
\label{fig:dens}}

\figcaption[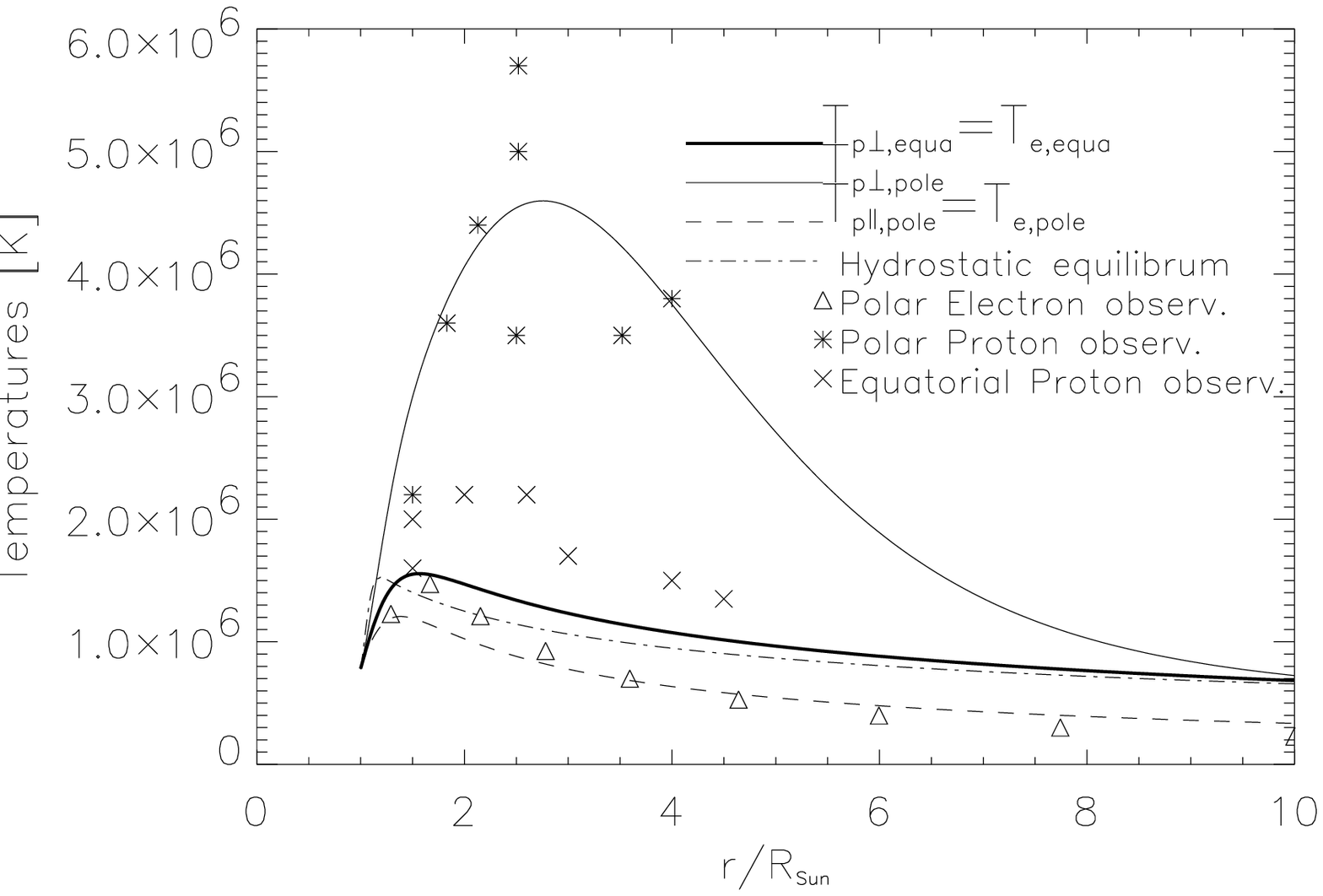]
{Temperature profiles derived from coronal observations,
for the pole and the equator.
\label{fig:tprof}}

\figcaption[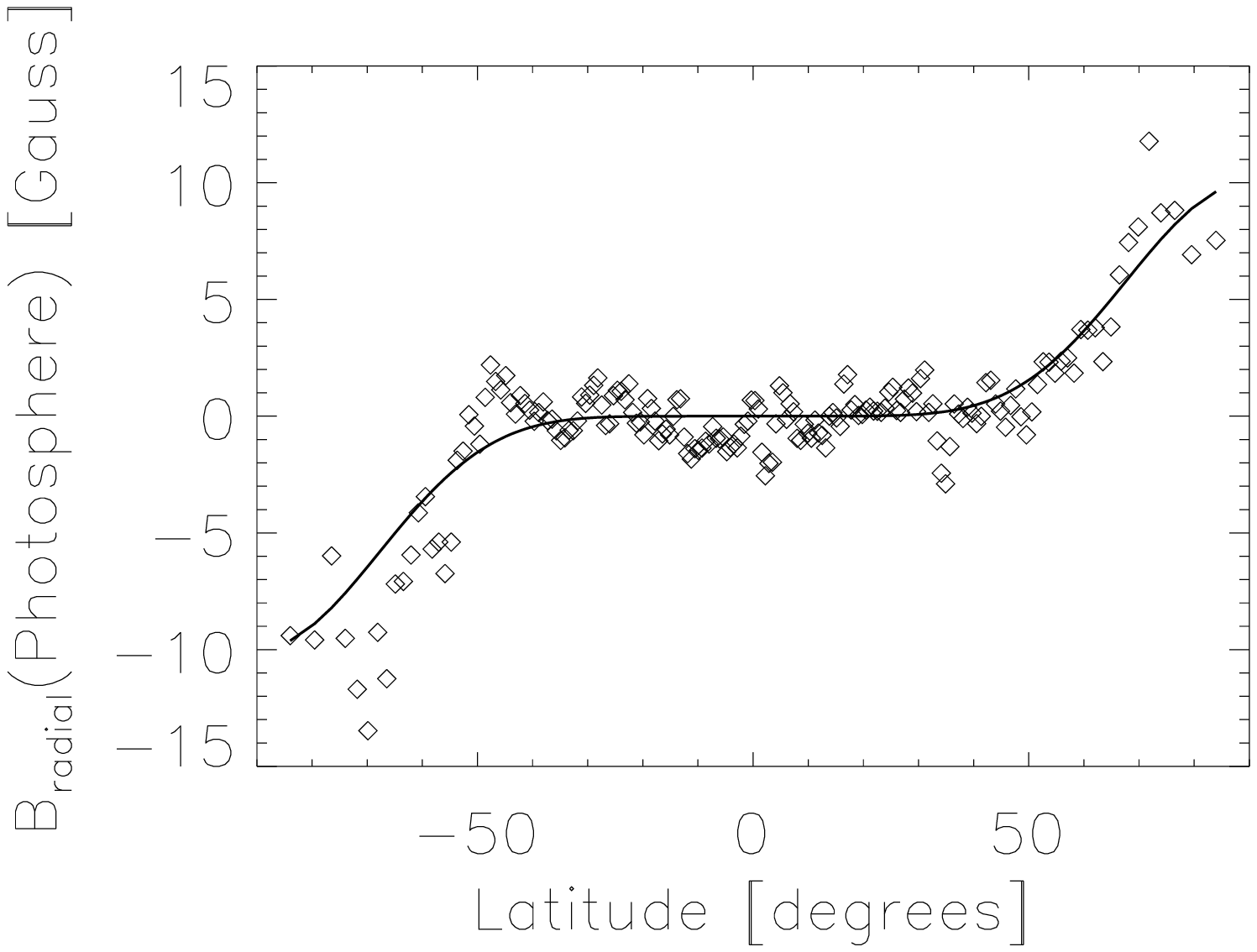]
{Photospheric magnetic field as function of latitude at the time of
the last cycle minimum (July 1996). The {\it diamonds} are
measurements from Kitt Peak synoptic maps. The {\it full} curve is a
fit to the data.
\label{fig:magn}}

\figcaption[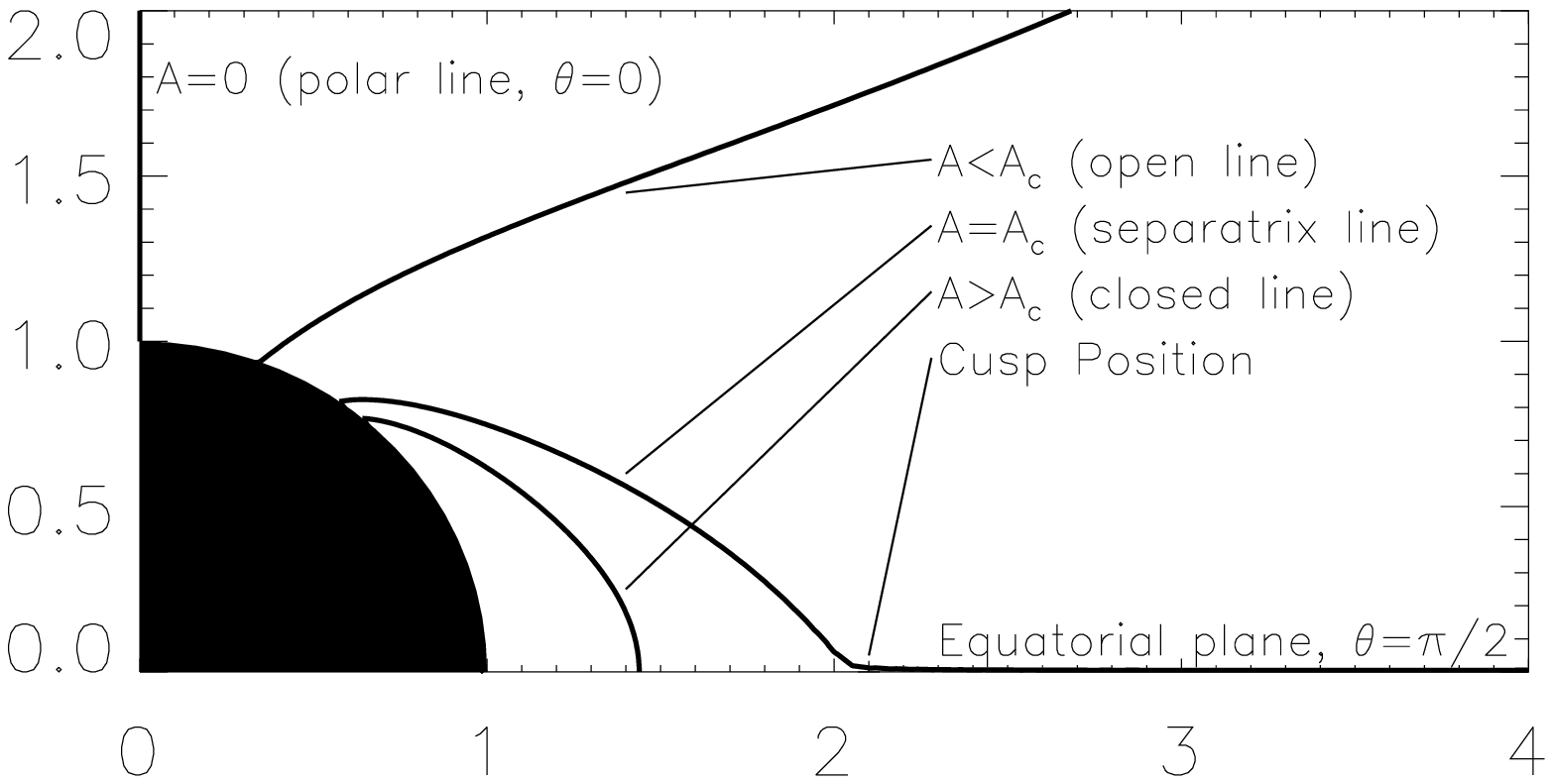]
{Sketch of the dipolar configuration.
\label{fig:sketch}}

\figcaption[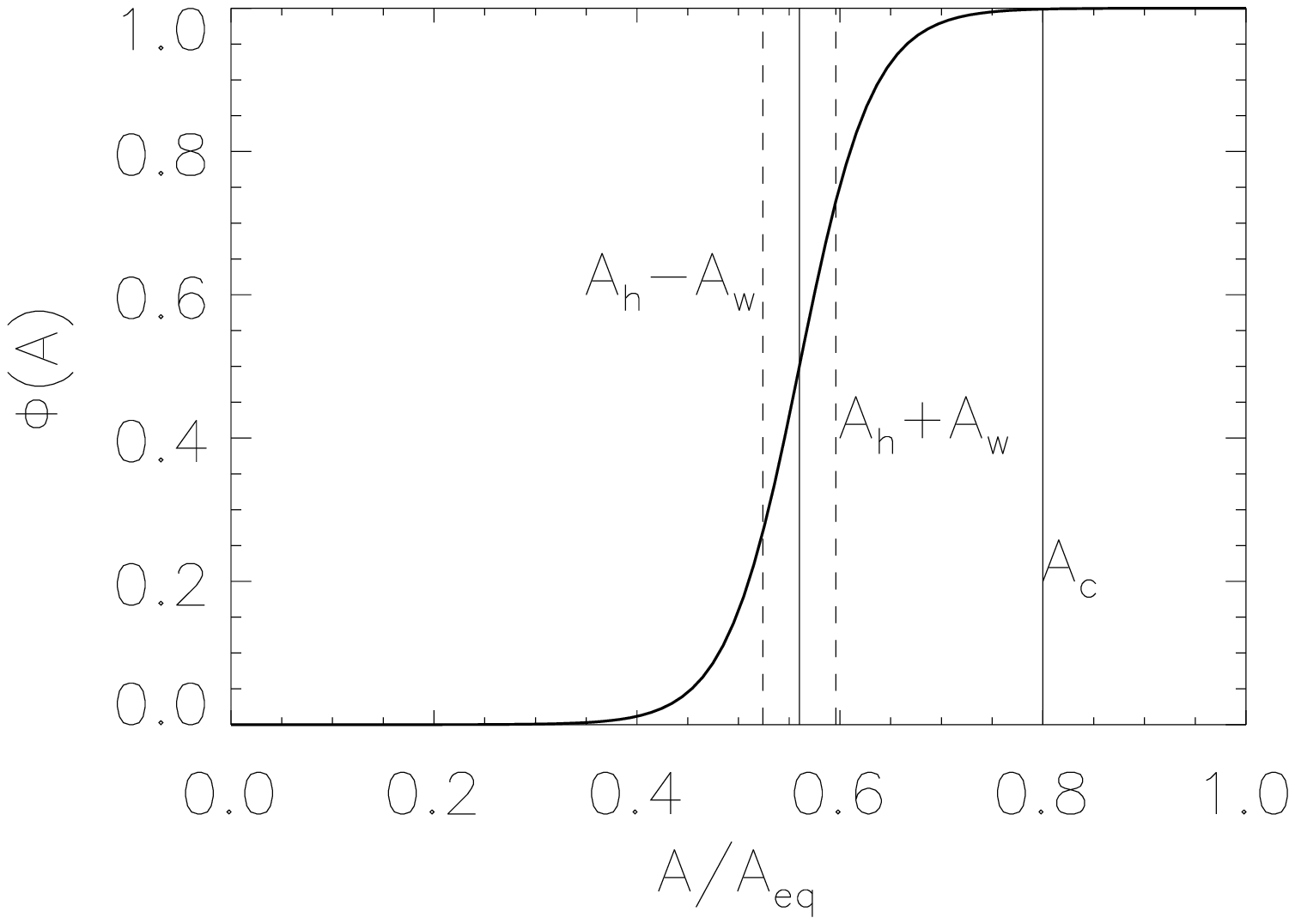]
{The function $\Phi(A)$ used for interpolating temperatures between
pole and equator.
\label{fig:phiA}}

\figcaption[f6.eps]
{Magnetic field and plasma pressure for the first iteration
({\it top}) and for the final model ({\it bottom}). The {\it thin}
curves are magnetic field lines (contours of $A$), and the
{\it thick} curves are contours of $\beta \equiv 8 \pi p_\perp /
B^2$. 
\label{fig:fld}}

\figcaption[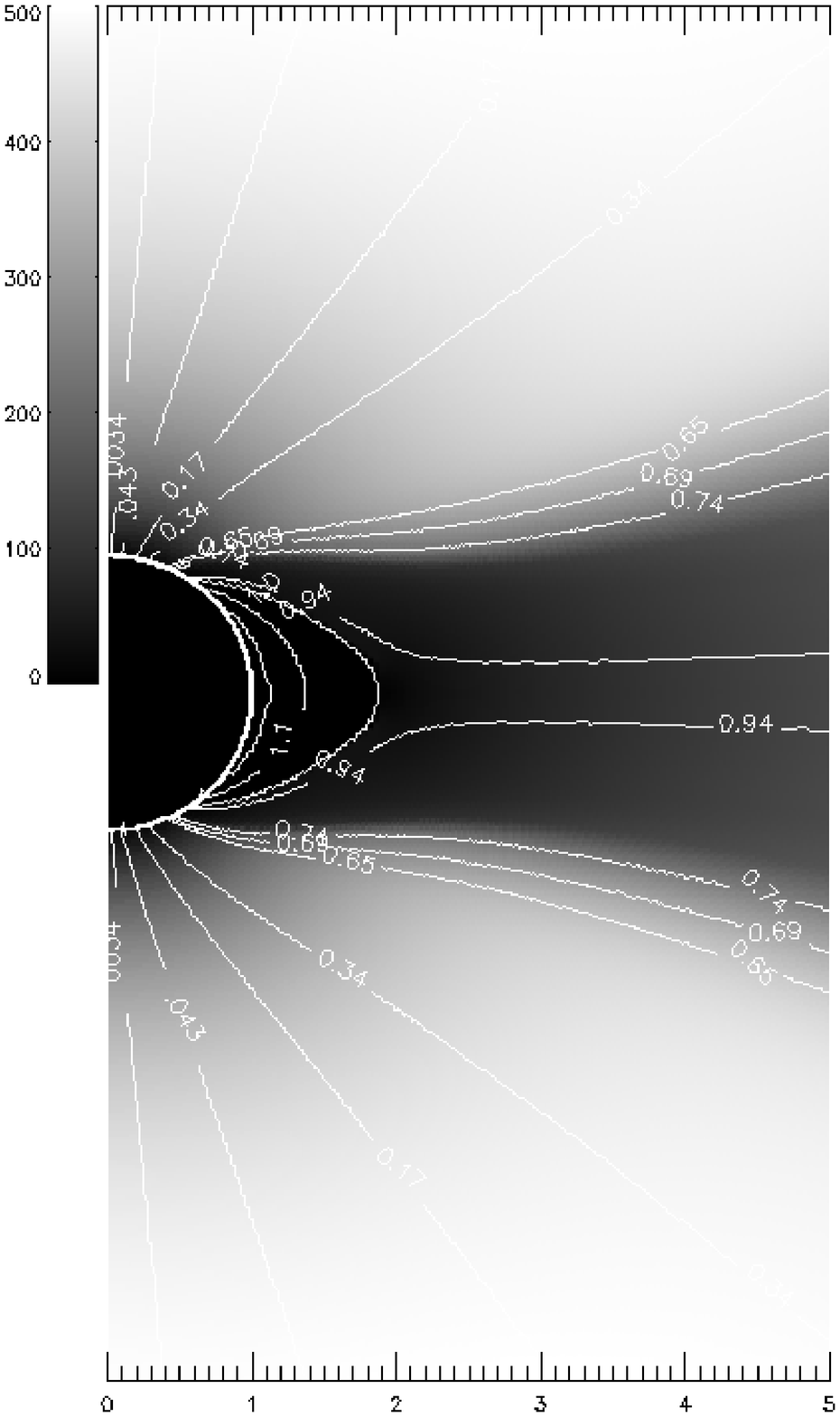]
{Outflow velocity in the meridional plane.
The brightness level increases with magnitude of the outflow speed,
$v$, as shown by the scale on the left side of the figure ($v$ in
km/s). The solar rotation axis is along the left axis of the plot, and
the solar surface is indicated by the {\it thick} semi-circle.
The {\it white} curves are selected field lines labeled with their
value of $A/A_c$.
\label{fig:velo}}

\figcaption[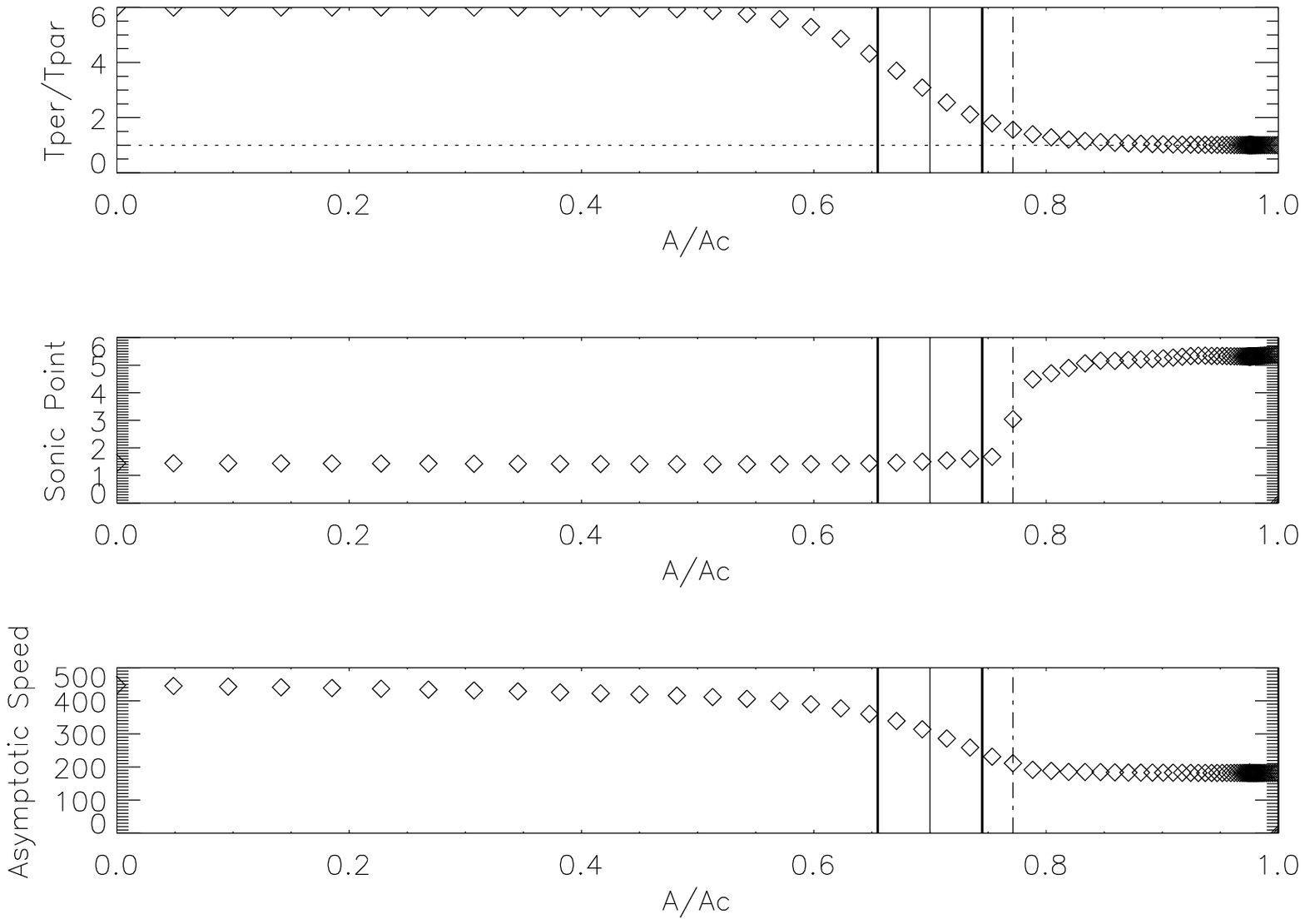]
{Various wind properties as function of field-line variable $A$.
{\it Top panel}: temperature anisotropy $(T_{p \perp} + T_e) /
(T_{p \parallel} + T_e)$ at $r = 3 \rsun$.
{\it Middle panel}: radius of sonic point where $v = c_\parallel$
(in $\rsun$).
{\it Bottom panel}: wind speed at $\rmax = 10 \rsun$ (in km/s).
The {\it thin solid} vertical line in each panel indicates the field
line $A_h$ where the latitudinal temperature gradient is largest,
and the {\it thick} vertical lines indicate the width of the transition
($A_h \pm A_w$). The {\it dot-dashed} vertical line indicates the field line
at which there is a sudden transition in the radius of the sonic
point.
\label{fig:across}}

\figcaption[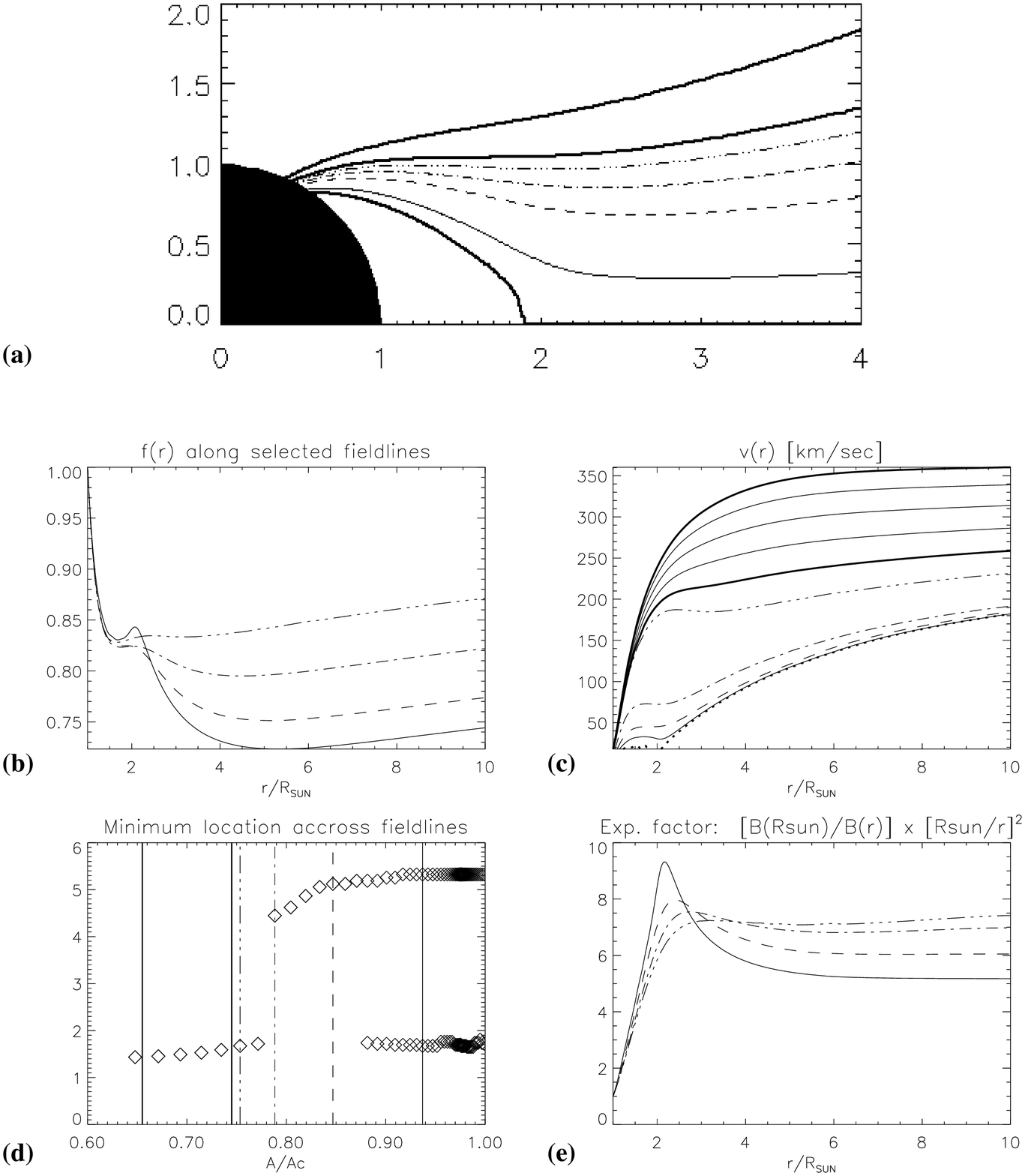]
{Transition between fast and slow wind.
(a) Shape of various field lines in the meridional plane.
(b) Functions $f(r)$ that appear on the RHS of the wind equation
(\ref{eq:wind2}) for different field lines.
(c) Outflow velocity $v(r)$ along field lines. The {\it dotted} curve
corresponds to the last open field line, $A = A_c$.
(d) Radial positions of the mimima of $f(r)$ as function of field-line
variable $A$ ({\it diamonds}). The vertical lines show the $A$-values
of the field lines shown in other panels.
(e) Non-radial expansion factors $f_{\rm exp}(r)$ for different field
lines.
The same line styles are used in all panels.
\label{fig:trans}}

\figcaption[f10.eps]
{Images of visible light polarization brightness ($pB$) as predicted
from the coronal model. Black curves are contours of $\log (pB /
pB_{max})$, where $pB_{max}$ is the value of $pB$ at the coronal
base. The white curves are selected field lines in the plane of the
sky (labeled by $A/A_c$). The dark semi-circle represents the solar
disk.
\label{fig:pB1}}

\figcaption[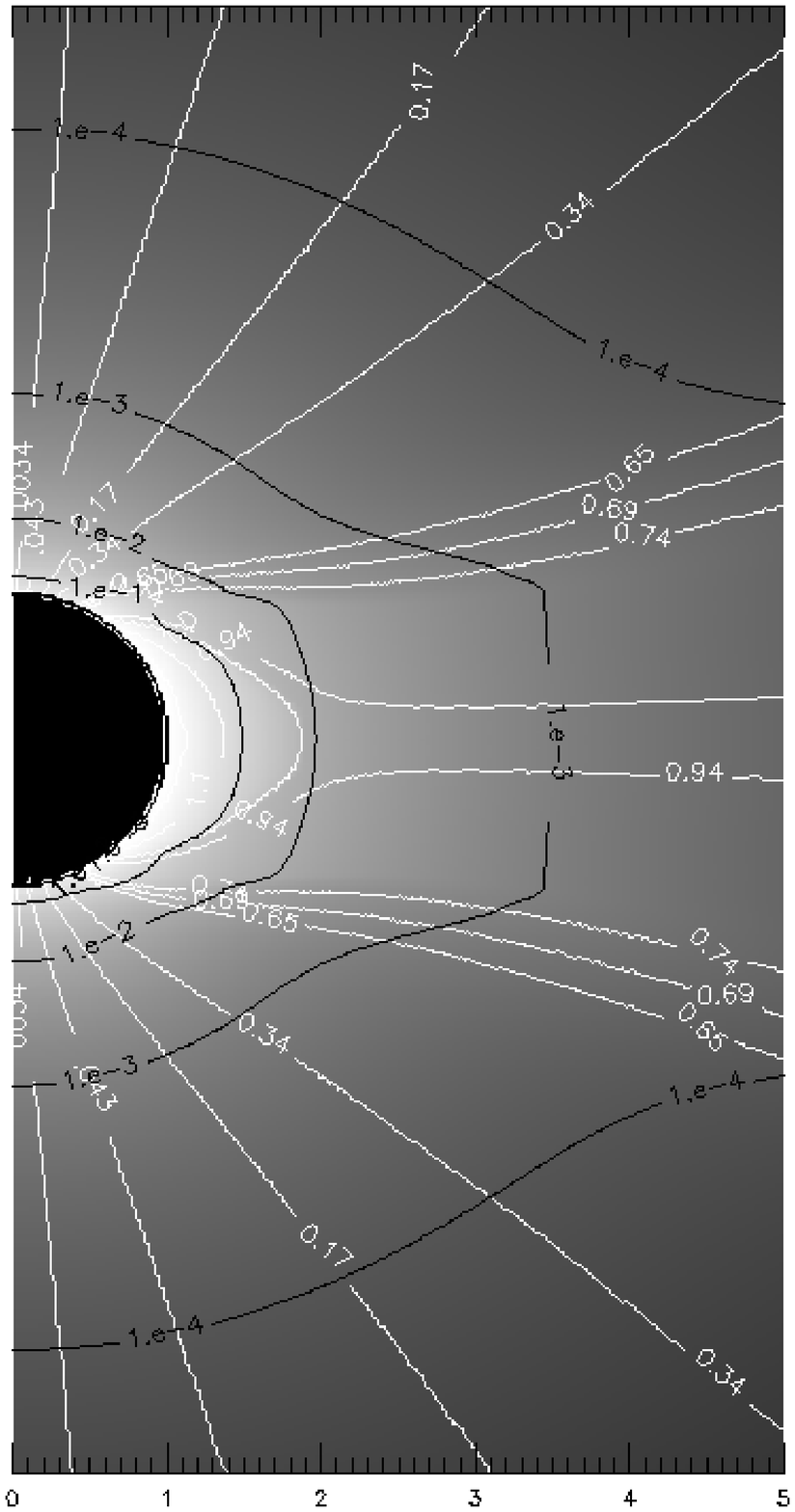]
{{\it Left panel}: Visible-light polarization brightness of the
corona, $pB/\overline{B}_\odot$, as function of radial distance
from sun center. The solid curves show the predicted equatorial
({\it thick}) and polar ({\it thin}) profiles for an edge-on view.
The dashed curve is the predicted polar profile for a view from
above the ecliptic plane ($\alpha = 30\deg$). The symbols are
measurements for the pole ({\it triangles}) and equator ({\it
diamonds}) from \citet{gibson99}, \citet{guhatha99}, and
\citet{cranmer99}.
{\it Right panel}: 
Predicted polarization brightness as function of co-latitude at
projected radii of 1.15, 1.5, and 2.5 $\rsun$. The symbols are
corresponding measurements from the above-cited references.
\label{fig:pB2}}

\figcaption[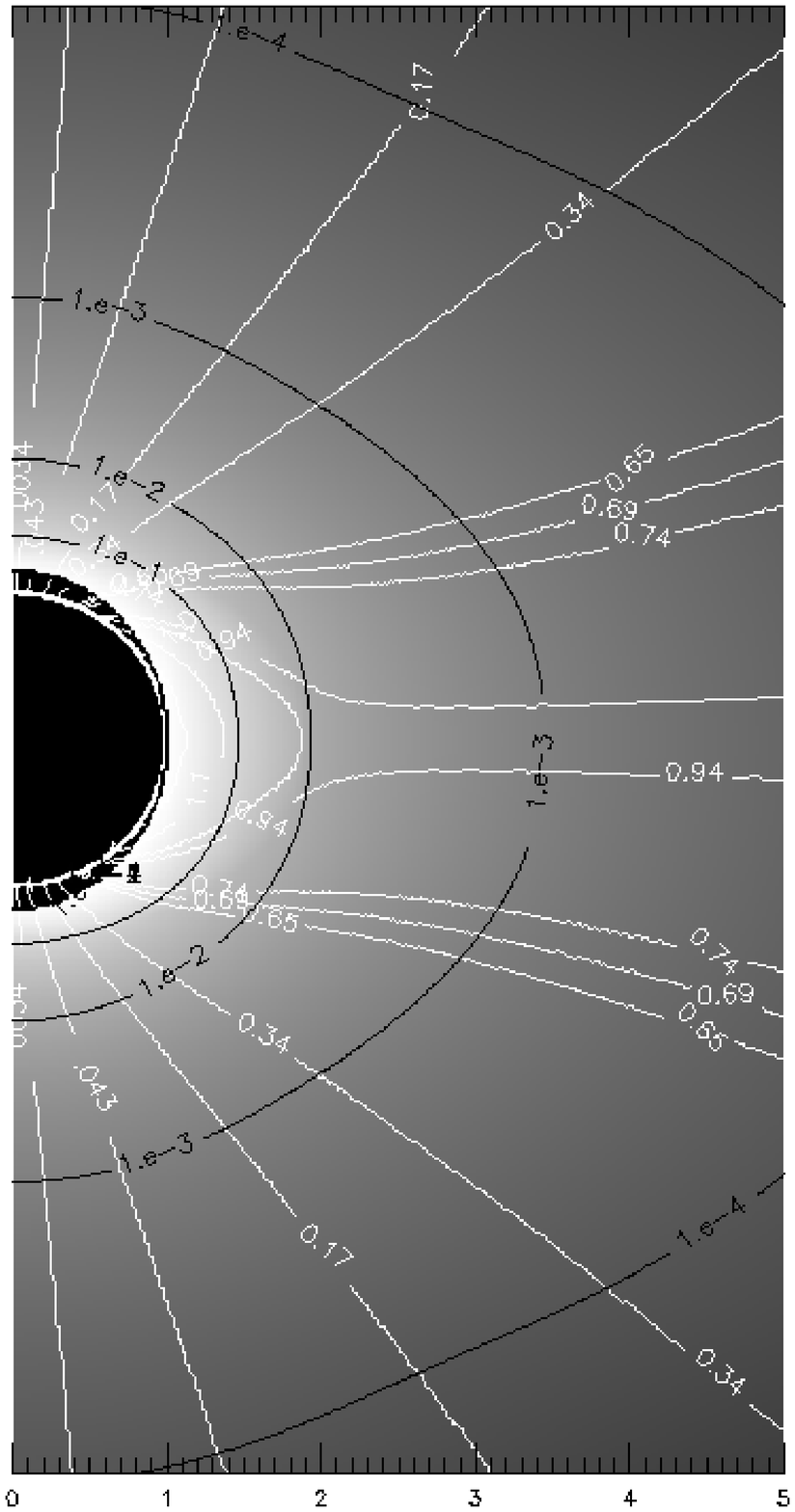]
{Images of Ly$\alpha$ intensity $I$ as predicted from the coronal
model. The black curves are contours of $\log (I / I_{max} )$,
where $I_{max}$ is the maximum value of $I$ at the coronal base.
The white curves are selected field lines in the plane of the sky
(labeled with $A/A_c$). The dark semi-circle represents the solar
disk.
\label{fig:Lya1}}

\figcaption[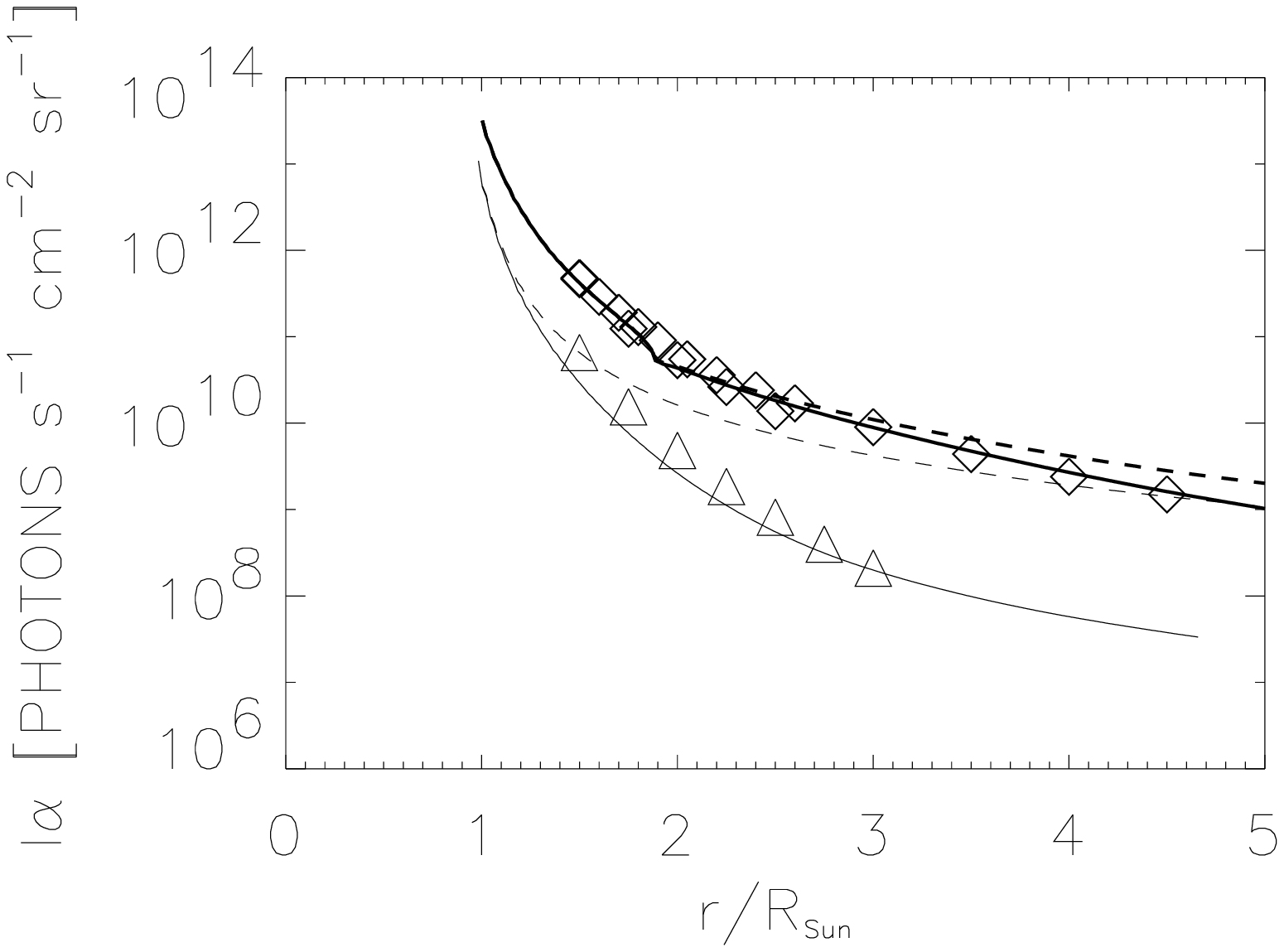]
{Radial profiles of Ly$\alpha$ intensity $I$. The solid curves
are the predicted intensity along the pole ({\it thin solid} curve)
and along the equator ({\it thick solid} curve) for the edge-on view
($\alpha = 0$). The {\it dashed} curves are the predicted profiles
without Doppler dimming [i.e., with $v_r = 0$ in equation
(\ref{eq:Lya})]. The {\it triangles} and {\it diamonds} are
intensity measurements from UVCS/{\it SOHO} for the pole and equator,
respectively (see text).
\label{fig:Lya2}}

%
%

\clearpage

\begin{deluxetable}{llll}
\tablewidth{0pt}
\tablecaption{Numerical values of the parameters for the temperature models.
\label{tab:temp_para}}
\tablehead{
\colhead{} & \colhead{$\Tpper$} & \colhead{$\Tppar$,$T_e$} &
\colhead{$\Tpper$,$\Tppar$,$T_e$} \\
\colhead{} & \colhead{at Pole} & \colhead{at Pole} & \colhead{at Equator} }
\startdata
$T_0$   & $8 \times 10^5$ K & $8 \times 10^5$ K & $8 \times 10^5$ K \\
$T_1$   & $3 \times 10^6$ K & 0. & 0. \\
$a$	   &0.		&0.	 &0.1	  \\
$b$	   &0.23	&0.47	 &0.33	  \\
$\alpha$ &0.7		&0.7	 &0.55    \\
$\beta$  &6.6		&6.6	 &6.6 	  \\
$\Delta r$ &1$\rsun$	& \nodata & \nodata
\enddata
\end{deluxetable}

%
%

\clearpage

\begin{figure}
\figurenum{1}
\epsscale{0.8}
\plotone{f1.eps}
\caption{}
\end{figure}

\begin{figure}
\figurenum{2}
\plotone{f2.eps}
\caption{}
\end{figure}

\begin{figure}
\figurenum{3}
\plotone{f3.eps}
\caption{}
\end{figure}

\begin{figure}
\figurenum{4}
\plotone{f4.eps}
\caption{}
\end{figure}

\begin{figure}
\figurenum{5}
\plotone{f5.eps}
\caption{}
\end{figure}

\begin{figure}
\figurenum{6}
\epsscale{1.25}
\plottwo{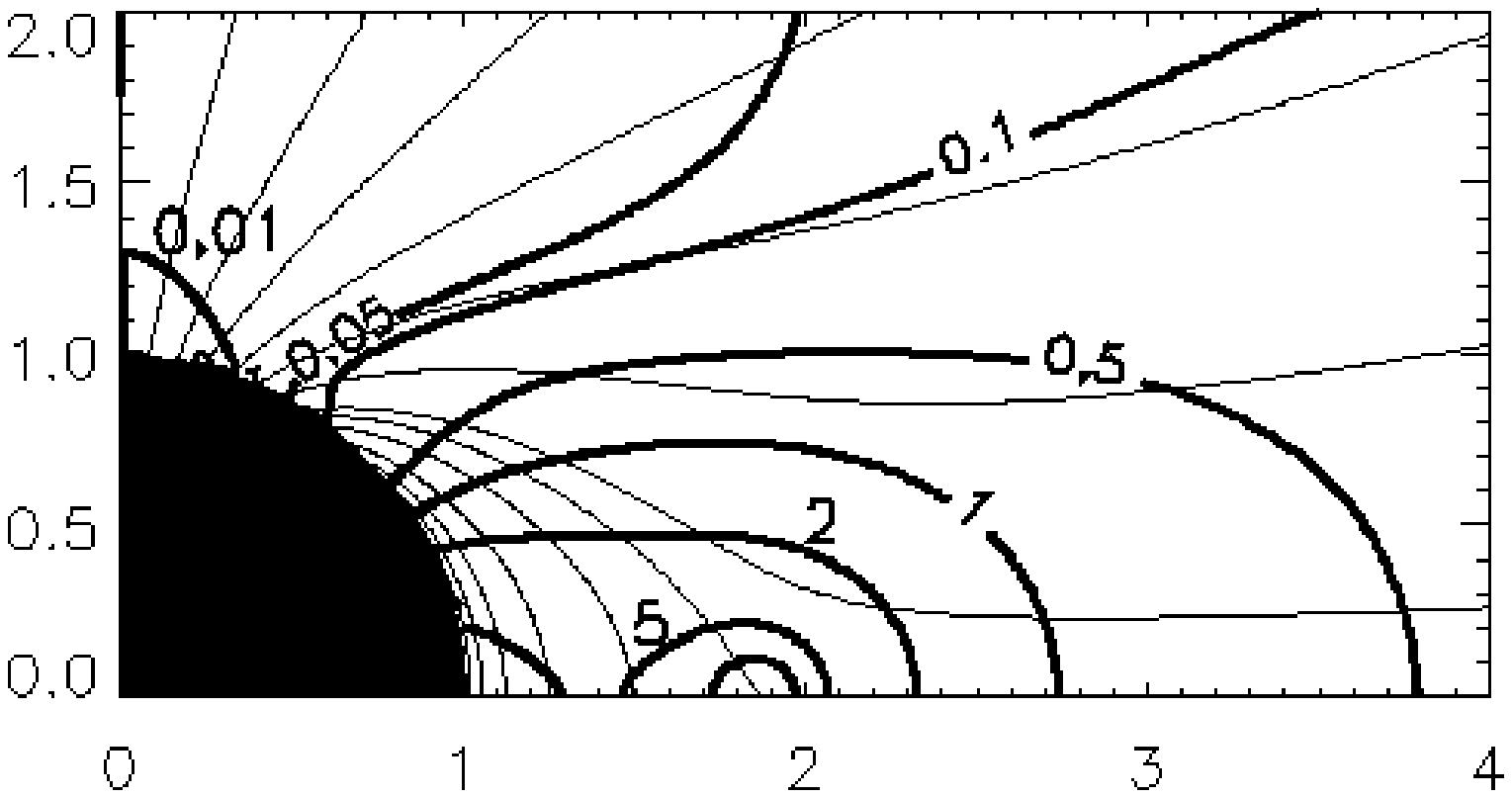}{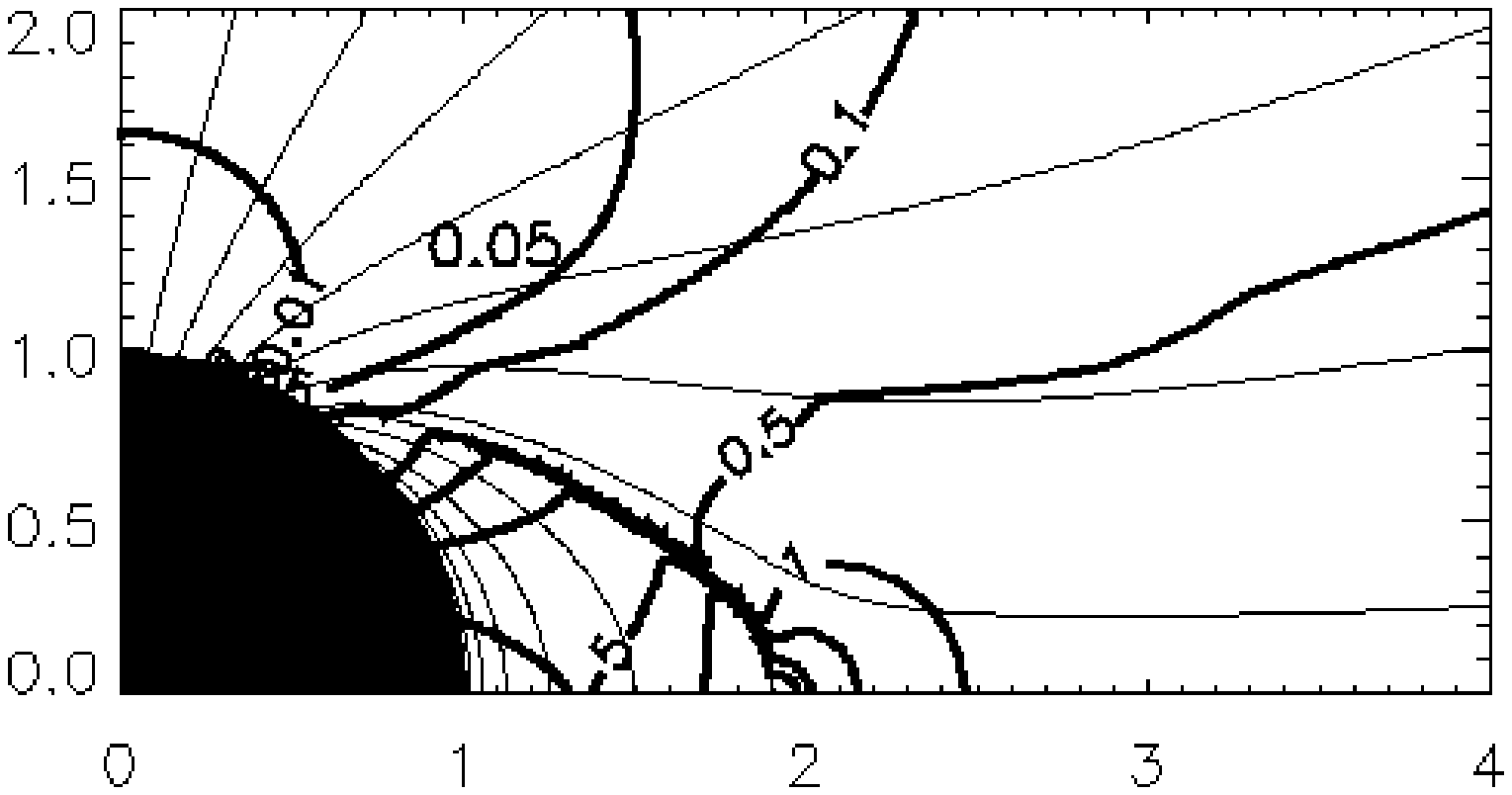}
\caption{}
\end{figure}

\begin{figure}
\figurenum{7}
\epsscale{.75}
\plotone{f7.eps}
\caption{}
\end{figure}

\begin{figure}
\figurenum{8}
\epsscale{1.}
\plotone{f8.eps}
\caption{}
\end{figure}

\begin{figure}
\figurenum{9}
\plotone{f9.eps}
\caption{}
\end{figure}

\begin{figure}
\figurenum{10}
\plottwo{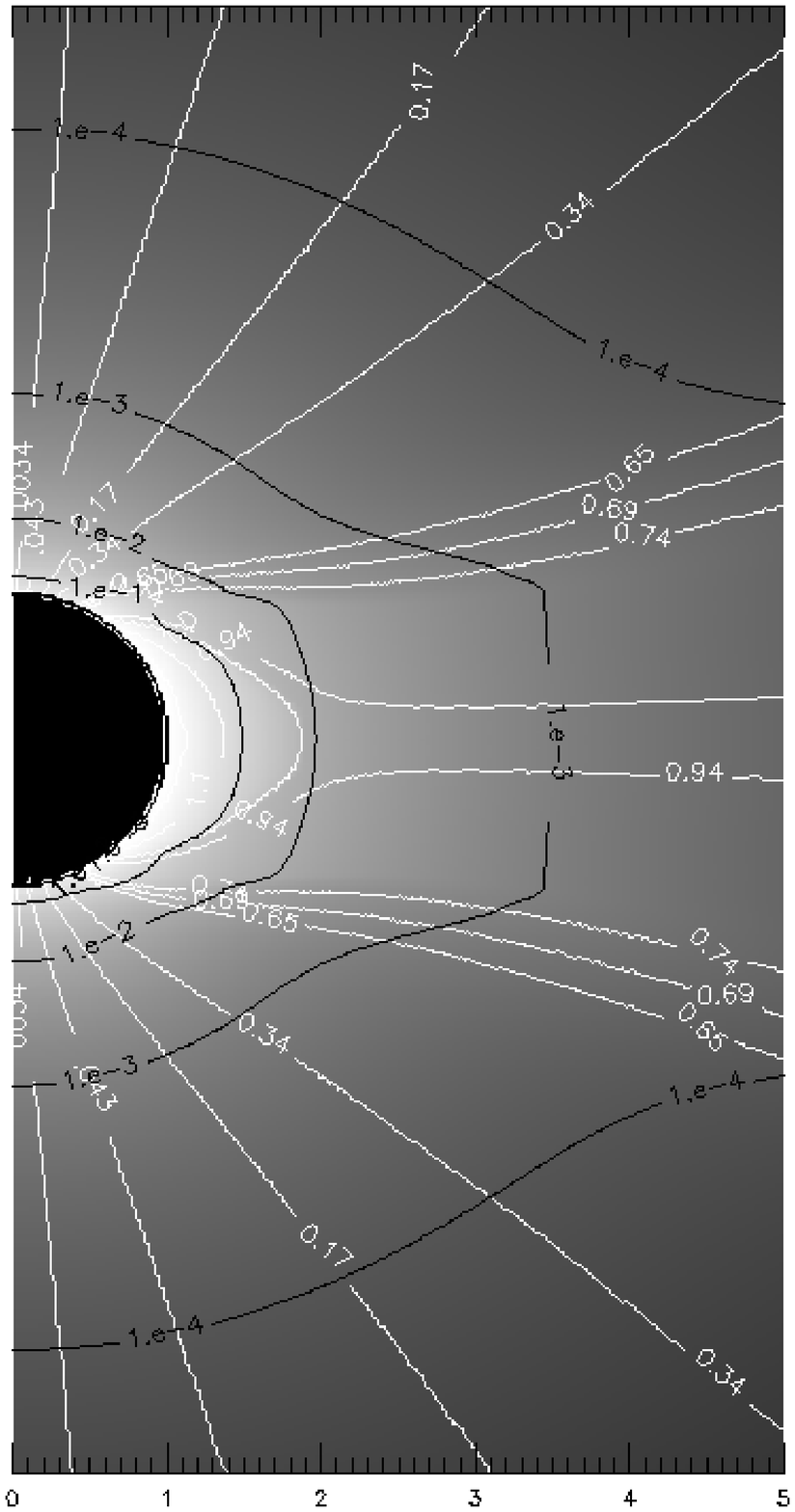}{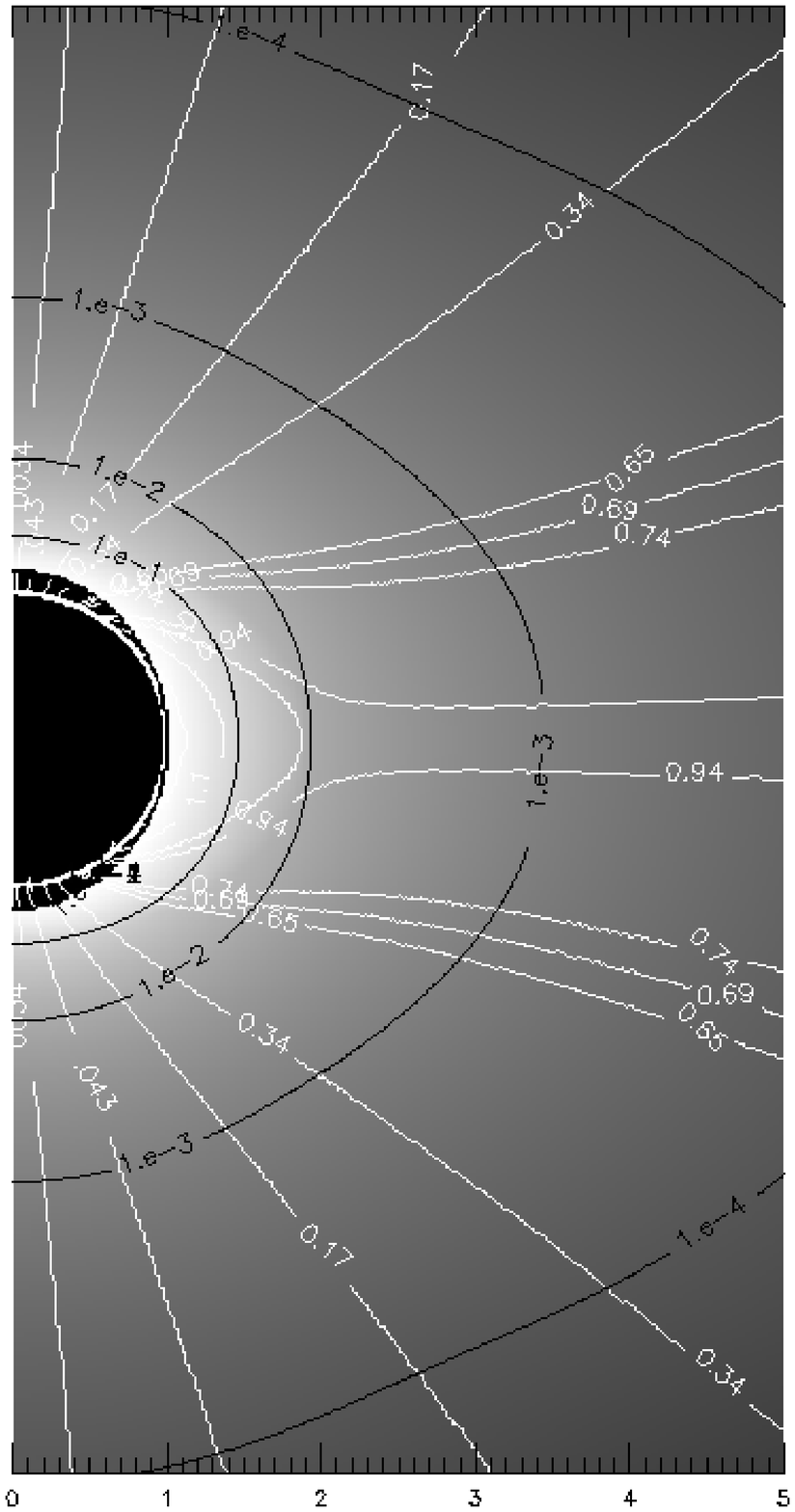}
\caption{}
\end{figure}

\begin{figure}
\figurenum{11}
\plottwo{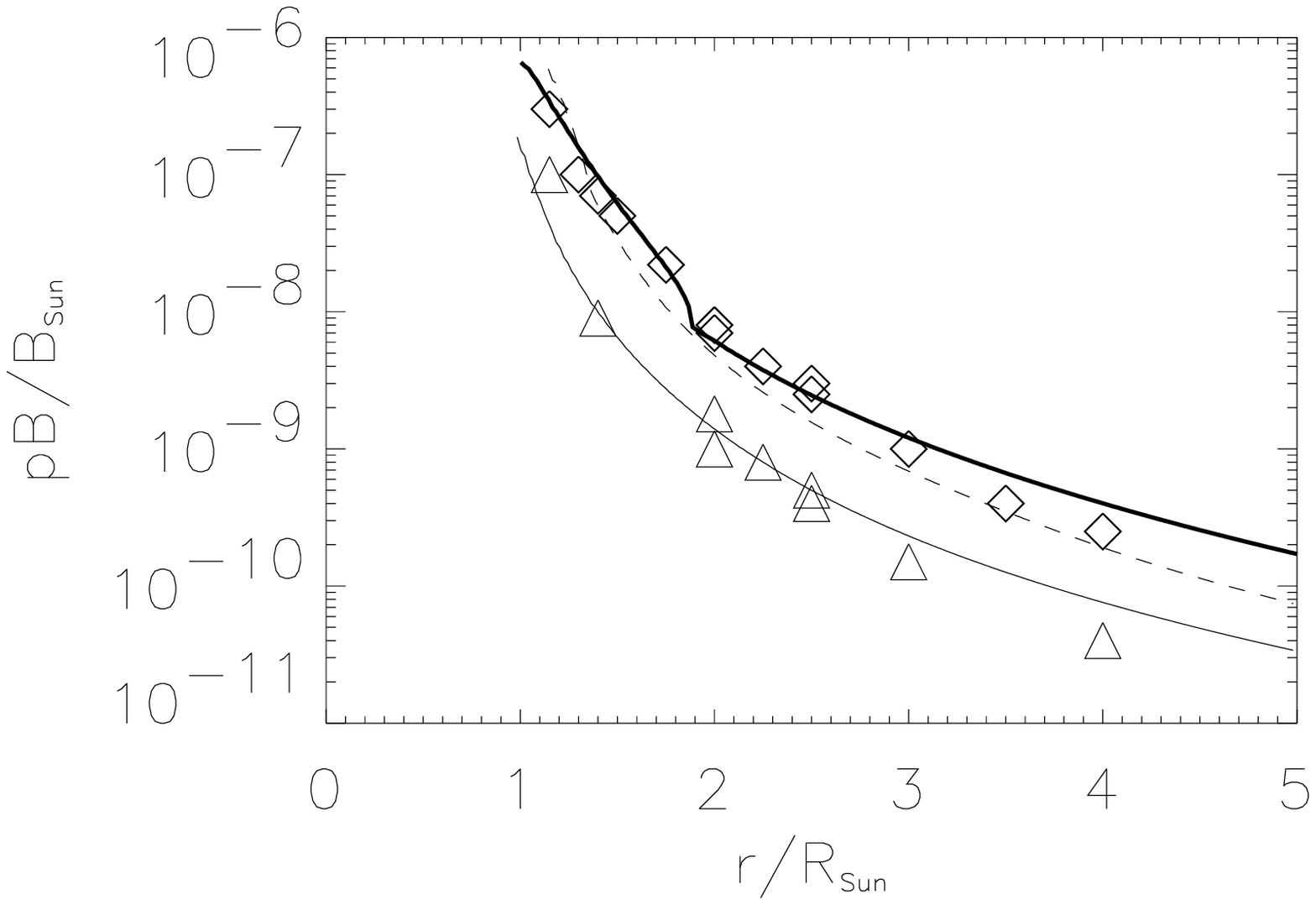}{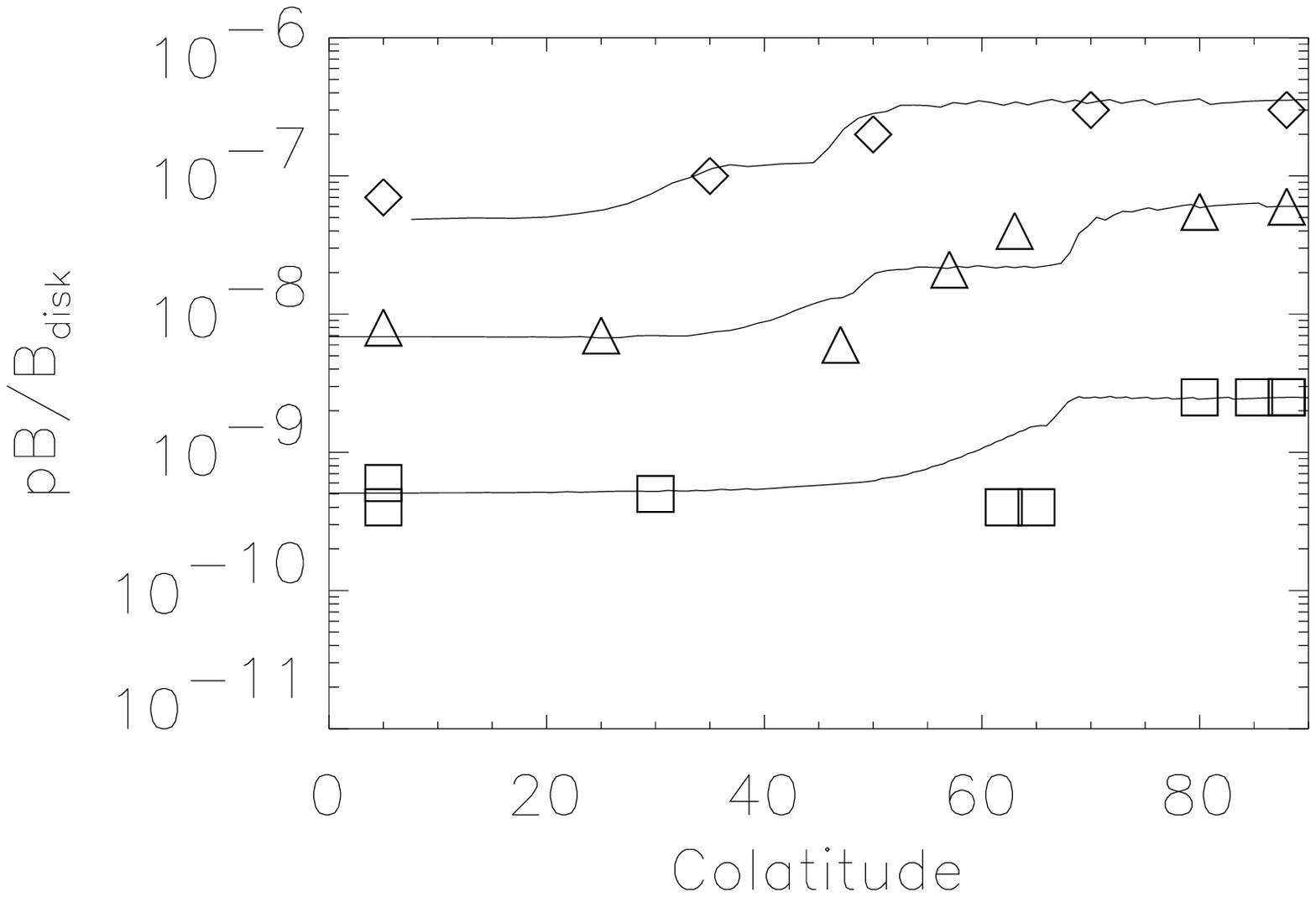}
\caption{}
\end{figure}

\begin{figure}
\figurenum{12}
\plottwo{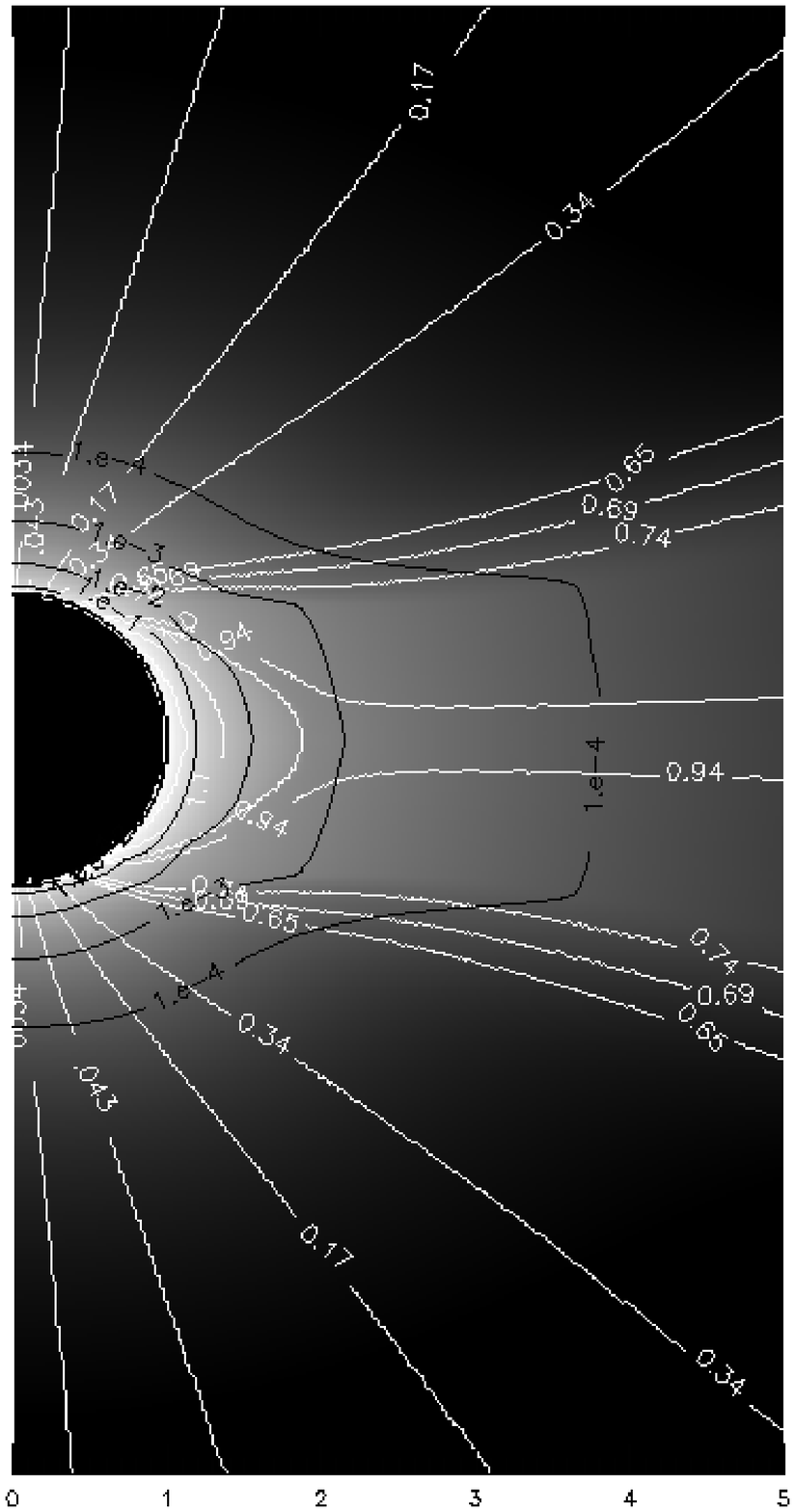}{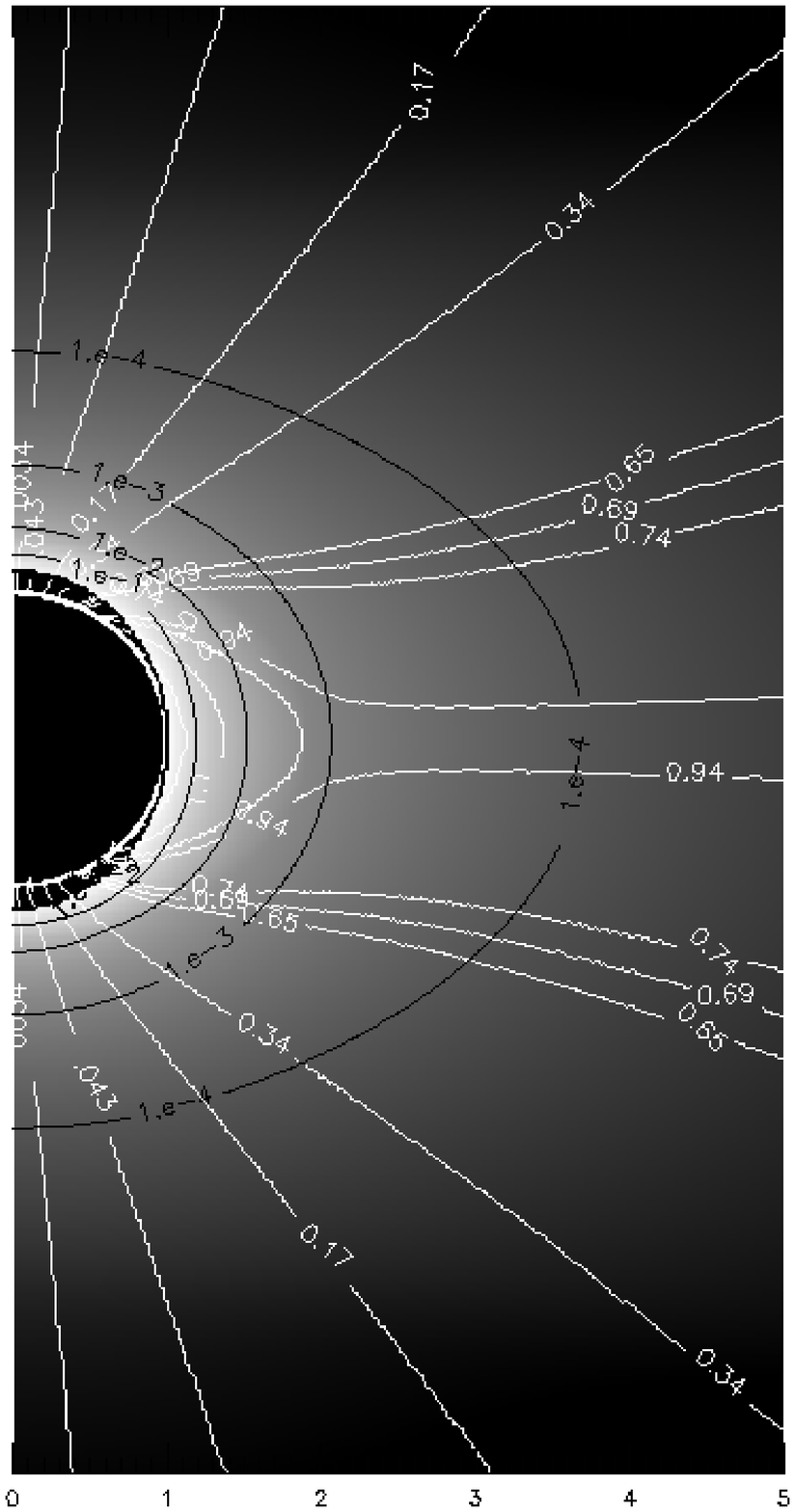}
\caption{}
\end{figure}

\begin{figure}
\figurenum{13}
\plotone{f13.eps}
\caption{}
\end{figure}


\begin{thebibliography}{DUM}

\bibitem[Allen, Habbal \& Hu(1998)]{allen98}
Allen, L. A., Habbal, S. R., \& Hu, Y. Q. 1998,
\jgr, 103, 6551

\bibitem[Allen(1973)]{allen73}
Allen, C. W. 1973, Astrophysical Quantities (London: Athlone)

\bibitem[Chen \& Hu(2001)]{chen2001}
Chen, Y., \& Hu, Y. Q. 2001, \solphys, 199, 371

\bibitem[Chen \& Hu(2002)]{chen2002}
Chen, Y., \& Hu, Y. Q. 2002, \apss, 282, 447

\bibitem[Cranmer(2002)]{cranmer2002}
Cranmer, S. R. 2002, in Proc. of 11th SOHO Symposium, From Solar Min
to Max: Half a Solar Cycle with SOHO, ed. A. Wilson, ESA SP-508
(Noordwijk: ESA), 361

\bibitem[Cranmer et al.(1999)]{cranmer99}
Cranmer, S. R., Kohl, J. L., Noci, G., Antonucci, E., Tondello, G.,
Huber, M. C. E., Strachan, L., et al. 1999, \apj, 511, 481

\bibitem[Cuperman et al.(1993)]{CB93}
Cuperman, S., Bruma, C., Detman, T., \& Dryer, M. 1993, \apj, 404, 356

\bibitem[Cuperman et al.(1990)]{CO90}
Cuperman, S., Ofman, L., \& Dryer M. 1990, \apj, 350, 846

\bibitem[Dobrzycka et al.(1999)]{danuta99}
Dobrzycka, D., Cranmer S. R., Panasyuk, A. V., Strachan L., \&
Kohl, J. L. 1999, \jgr, 104, 9791

\bibitem[Dusenbery \& Hollweg(1981)]{dusen81}
Dusenbery, P. B., \& Hollweg, J. V. 1981, \jgr, 86, 153

\bibitem[Frazin, Cranmer \& Kohl(2003)]{frazin2003}
Frazin, R. A., Cranmer, S. R., and Kohl, J. L. 2003, ApJ, submitted

\bibitem[Frazin(2002)]{frazin2002b}
Frazin, R. A. 2002, Empirical Constraints on O5+ Outflows and Velocity
Distributions in a Solar Minimum Coronal Streamer, Ph.D. Thesis,
The University of Illinois at Urbana-Champaign

\bibitem[Gibson et al.(1999)]{gibson99}
Gibson, S. E., Fludra, A., Bagenal, F., Biesecker, D., Del Zanna, G.,
\& Bromage, B. 1999, \jgr, 104, 9691

\bibitem[Gibson(2001)]{gibson01}
Gibson, S. E. 2001, \ssr, 97, 69

\bibitem[Guhathakurta et al.(1999)]{guhatha99}
Guhathakurta, M., Fludra, A., Gibson, S. E., Biesecker, D., \&
Fisher, R. 1999, \jgr, 104, 9801

\bibitem[Guhathakurta, Holzer \& MacQueen(1996)]{guhatha96}
Guhathakurta, M., Holzer, T. E., \& MacQueen, R.M. 1996, \apj, 458,
817

\bibitem[Habbal et al.(1997)]{habbal97}
Habbal, S. R., Woo, R., Fineschi, S., O'Neal, R., Kohl, J., Noci, G.,
\& Korendyke, C. 1997, \apj, 489, L103

\bibitem[Hu \& Habbal(1999)]{hu+habbal99}
Hu, Y. Q., \& Habbal, S. R. 1999, \jgr, 104, 17045

\bibitem[Hummer(1994)]{hummer94}
Hummer, D. G. 1994, \mnras, 268, 109

\bibitem[Isenberg \& Hollweg(1983)]{isen83}
Isenberg, P. A., \& Hollweg, J. V. 1983, \jgr, 88, 3293

\bibitem[Ko et al.(1997)]{ko97}
Ko, Y.-K., Fisk, L. A., Geiss, J., Gloeckler, G., Guhathakurta,
M. 1997, \solphys, 171, 345

\bibitem[Kohl et al.(1995)]{kohl95}
Kohl, J. L., Esser, R., Gardner, L. D., Habbal, S., Daigneau, P. S.,
Dennis, E. F., Nystrom, G. U., et al. 1995, \solphys, 162, 313

\bibitem[Kohl et al.(1997)]{kohl97}
Kohl, J. L., Noci, G., Antonucci, E., Tondello, G., Huber, M. C. E.,
Gardner, L. D., Nicolosi, P., Strachan, L., et al. 1997, \solphys,
175, 613

\bibitem[Kohl et al.(1998)]{kohl98}
Kohl, J. L., Noci, G., Antonucci, E., Tondello, G., Huber, M. C. E.,
Cranmer, S. R., Strachan, L., et al. 1998, \apjl, 501, L127

\bibitem[Kohl, Strachan \& Gardner(1996)]{KSG96}
Kohl, J. L., Strachan, L., \& Gardner, L. D. 1996, \apjl, 465, L141

\bibitem[Landi et al.(2001)]{landi01}
Landi, E., Doron, R., Feldman, U., \& Doschek, G. A. 2001., \apj, 556,
912

\bibitem[Li et al.(1998)]{li98}
Li, J., Raymond, J. C., Acton, L. W., Kohl, J. L., Romoli, M., Noci,
G., \& Naletto, G. 1998, \apjl, 506, 431L

\bibitem[Linker, Van Hoven \& Schnack(1990)]{LH90}
Linker, J. A., Van Hoven, G., and Schnack, D. D. 1990, \jgr 17, 2281

\bibitem[Lionello, Linker \& Mikic(2001)]{lionello2001}
Lionello, R., Linker, J. A., \& Mikic, Z. 2001, \apj, 546, 542

\bibitem[Low(1986)]{low86}
Low, B. C. 1986, \apj, 310, 953

\bibitem[Marsh et al.(1982)]{marsh82}
Marsh, E., Muhlhauser, K.-H., Rosenbauer, H., Schwenn, R., \&
Neubauer, F. M. 1982, \jgr, 86, 9199

\bibitem[Munro \& Jackson(1977)]{mj77}
Munro, R. H., \& Jackson, B. V. 1977, \apj, 213, 874

\bibitem[Newkirk(1967)]{newkirk67}
Newkirk, G. A., Jr. 1967, \araa, 5, 213

\bibitem[Noci et al.(1997)]{noci97}
Noci, G., Kohl, J. L., Antonucci, E., Tondello, G., Huber, M. C. E.,
Fineschi, S., Gardner, L. D., et al. 1997, in Proc. Fifth {\it SOHO}
Workshop: The Corona and Solar Wind Near Minimum Activity,
ed. A. Wilson, ESA SP-404 (Noordwijk: ESA), 75

\bibitem[Noci, Kohl \& Withbroe(1987)]{noci87}
Noci, G., Kohl, J. L., \& Withbroe, G. L. 1987, \apj, 315, 706

\bibitem[Olsen \& Leer(1996)]{olsen96}
Olsen, E. L., \& Leer, E. 1996, \apj, 462, 982

\bibitem[Olsen, Leer \& Holzer(1994)]{olsen94}
Olsen, E. L., Leer, E., \& Holzer, T. 1994, \apj, 420, 913

\bibitem[Phillips et al.(1995)]{phillips95}
Phillips, J. L., Bame, S. J., Barnes, A., Barraclough, B. L., Feldman,
W. C., Goldstein, B. E., Gosling, J. T., et al. 1995, \grl, 22, 3301

\bibitem[Pilipp et al.(1987)]{pilipp87}
Pilipp, W. G., Muehlhaeuser, K.-H., Miggenrieder, H., Montgomery, M. D., 
Rosenbauer, H. 1987, \jgr, 92, 1075

\bibitem[Pneuman \& Kopp(1971)]{PK71}
Pneuman, G., \& Kopp, R. A. 1971, \solphys, 18, 258

\bibitem[Press, et al.(1992)]{NR}
Press, W. H., Teucholsky, S. A., Vetterling, W. T., \& Flannery, B. P.
1992, Numerical Recipes in Fortran: The Art of Scientific
Computing, 2nd Edition (Cambridge University Press), 413

\bibitem[Raymond et al.(1997)]{raymond97}
Raymond, J. C., Kohl, J. L., Noci, G., Antonucci, E., Tondello, G.,
Huber, M. C. E., Gardner, L. D., Nicolosi, P., et al. 1997, \solphys,
175, 645

\bibitem[Riley, Linker \& Mikic(2001)]{riley2001}
Riley, P., Linker, J. A., \& Mikic, Z. 2001, \jgr, 106, 15889

\bibitem[Saito et al.(1977)]{saito77}
Saito, K., Poland, A. I., \& Munro., R. H. 1977, \solphys, 55, 121

\bibitem[Scholz \& Walters(1991)]{SW91}
Scholz, T. T., \& Walters, H. R. J. 1991, \apjs, 380, 302

\bibitem[Sheeley et al.(1997)]{sheeley97}
Sheeley, N. R., Jr., Wang, Y.-M., Hawley, S. H., Brueckner, G. E.,
Dere, K. P., Howard, R. A., Koomen, M. J., et al. 1997, \apj, 484, 472

\bibitem[Sittler \& Guhathakurta(1999)]{sittler99}
Sittler, E. C., Jr., \& Guhathakurta, M. 1999, \apj, 523, 812

\bibitem[Steinolfson, Suess \& Wu(1982)]{SS82}
Steinolfson, R. S., Suess, S. T., \& Wu, S. T. 1982, \apj, 255, 730

\bibitem[Strachan et al.(2002)]{strachan2002}
Strachan, L., Suleiman, R., Panasyuk, A. V., Biesecker, D. A., \&
Kohl, J. L. 2002, \apj, 571, 1008

\bibitem[Suess \& Nerney(1999)]{SuNe99}
Suess, S. T., \& Nerney, S. F. 1999, in Proc. 9th European Meeting
on Solar Physics, Magnetic Fields and Solar Processes, ed. A. Wilson,
ESA SP-44 (Noordwijk: ESA), 1101

\bibitem[Suess \& Nerney(2002)]{SuNe02}
Suess, S. T., \& Nerney, S. F. 2002, \apj, 565, 1275

\bibitem[Suess, Gary \& Nerney(1999)]{SuGaNe99}
Suess, S. T., Gary, G. A., \& Nerney, S. F. 1999b, in Proc. of the
Ninth International Solar Wind Conference, eds. S. R. Habbal,
R. Esser, J. V. Hollweg, \& P. A. Isenberg, AIP Conf. Proc., Vol. 471,
247

\bibitem[Suess et al.(1999)]{Su99}
Suess, S. T., Poletto, G., Corti, G., Simnett, G., Noci, G.,
Romoli, R., Kohl, J., \& Goldstein, B. 1999, \ssr, 87, 319

\bibitem[Suess et al.(1996)]{Su96}
Suess, S. T., Wang, A.-H., \& Wu, S. T. 1996, \jgr, 101, 19957

\bibitem[Tu \& Marsch(1997)]{tu+marsch97}
Tu, C.-Y., \& Marsch, E. 1997, \solphys, 171, 363

\bibitem[van de Hulst(1950)]{hulst50}
van de Hulst, H. C. 1950, Bull. Astron. Inst. Neth., 11, 135

\bibitem[V\'asquez, Raymond \& van Ballegooijen(1999)]{Vas99a}
V\'asquez, A. M., Raymond, J. C., \& van Ballegooijen, A. A. 1999,
\ssr, 87, 335

\bibitem[V\'asquez, van Ballegooijen \& Raymond(1999)]{Vas99b}
V\'asquez, A. M., van Ballegooijen, A. A., \& Raymond, J. C. 1999, 
in Proc. of the Ninth International Solar Wind Conference,
eds. S. R. Habbal, R. Esser, J. V. Hollweg, \& P. A. Isenberg, AIP
Conf. Proc., Vol. 471, 243

\bibitem[Wang A. H. et al.(1993)]{WWSP93}
Wang, A. H., Wu, S. T., Suess, S. T., \& Poletto, G. 1993, \solphys,
147, 55

\bibitem[Wang A. H. et al.(1998a)]{WWSP98}
Wang, A. H., Wu, S. T., Suess, S. T., \& Poletto, G. 1998a, \jgr, 103,
1913

\bibitem[Wang Y.-M.(1994)]{wang94}
Wang, Y.-M. 1994, \apjl, 437, L67

\bibitem[Wang \& Sheeley(1990)]{wang90}
Wang, Y.-M., \& Sheeley, N. R., Jr. 1990, \apj, 355, 726

\bibitem[Wang Y.-M. et al.(1997a)]{wang97a}
Wang, Y.-M., Sheeley, N. R., Jr., Howard, R. A., Kraemer, J. R., Rich,
N. B., Andrews, M. D., Brueckner, G. E., et al. 1997a, \apj, 485, 875

\bibitem[Wang Y.-M. et al.(1997b)]{wang97b}
Wang, Y.-M., Sheeley, N. R., Jr., Phillips, J. L., \& Goldstein,
B. E. 1997b, \apjl, 488, L51

\bibitem[Wang Y.-M. et al.(1998b)]{wang98}
Wang, Y.-M., Sheeley, N. R., Jr., Walters, J. H., Brueckner, G. E.,
Howard, R. A., Michels, D. J., Lamy, P. L., Schwenn, R., Simnett,
G. M. 1998b, \apjl, 498, L165

\bibitem[Washimi, Yoshino \& Ogino(1987)]{WY87}
Washimi, H., Yoshino, Y. and Ogino, T. 1987, \jgr, 14, 487

\bibitem[Wilhelm et al.(1998)]{wilhelm98}
Wilhelm, K., Marsch, E., Dwivedi, B. N., Hassler, D. M., Lemaire, Ph.,
Gabriel, A. H., Huber, M. C. E. 1998, \apj, 500, 1023

\bibitem[Withbroe et al.(1982)]{wk82}
Withbroe, G. L., Kohl, J. L., Weiser, H., \& Munro, R. H. 1982, \ssr,
33, 17

\bibitem[Yeh \& Pneuman(1977)]{YP77}
Yeh, T., \& Pneuman, G. W. 1977, \solphys, 54, 419

\end{thebibliography}
\end{document}